\documentclass[aps,pre,reprint]{revtex4-2}
\usepackage{amsmath}
\usepackage{amssymb}
\usepackage{graphicx}

\begin{document}

\title{Convective Space-Time Chaos as a Dynamical Model of Deterministic and Stochastic Turbulence}

\author{Arkady Pikovsky}

\affiliation{Institute for Physics and Astronomy, University of Potsdam, Potsdam, Germany}



\begin{abstract}
Recently, a concept of deterministic and stochastic turbulence has been introduced based on experiments with a boundary layer. In these experiments, the flow was driven with controlled random perturbation; in addition, natural ambient noise was also present.  Deterministic property manifested itself as repeatability of turbulence patterns induced by identical random perturbations at the inlet (deterministic turbulence).  A stochastic non-identical component originating from natural noise grows and eventually dominates the flow further downstream (stochastic turbulence). We argue that these properties can be explained by exploring the concept of convective space-time chaos, where secondary perturbations on top of a chaotic state grow but move away in the laboratory reference frame. We illustrate this with two simple models of convective space-time chaos, one is a partial differential equation describing waves on a film flowing down a plate, and the other is a set of unidirectionally coupled ordinary differential equations. To prove convective space-time chaos, we calculate the profiles of the convective Lyapunov exponent. The repeatability of the turbulent field in different identical experimental runs corresponds to the reliability of stable dynamical systems in response to random forcing. The onset of the stochastic component is quantified with the spatial Lyapunov exponent. We demonstrate how an effective randomization of the field is observed when the driving is quasiperiodic. Furthermore, we discuss space-time duality, which links sensitivity to boundary conditions in the convective space-time chaos to the usual sensitivity to initial conditions in a standard chaotic regime.
\end{abstract}
\maketitle
\section{Introduction}
\label{sec:intro}
Since the seminal papers of Lorenz~\cite{Lorenz-63} and Ruelle and Takens~\cite{Ruelle-Takens-71}, chaos serves as a basis for the explanation of the deterministic unpredictability of turbulent flows. Although for flows in constrained geometries already simple low-dimensional chaotic systems provide a good description, for more developed turbulence, one needs rather advanced dynamical models of space-time chaos~\cite{bohr1998dynamical}. Albeit such models do not reproduce the detailed features of turbulence, which are only available through thorough direct numerical simulations, they serve as useful conceptual tools for understanding qualitative features such as predictability, controllability, cascades across scales, etc.

\begin{figure}[!htb]
\centering
\includegraphics[width=0.48\columnwidth]{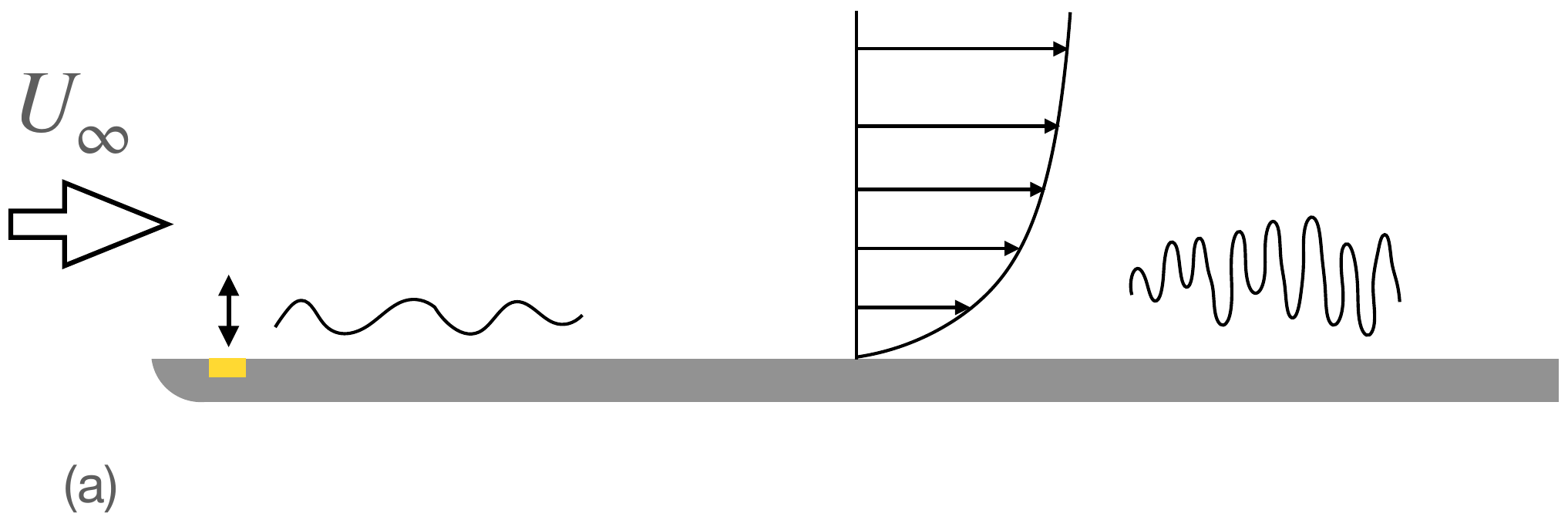}\hfill
\includegraphics[width=0.48\columnwidth]{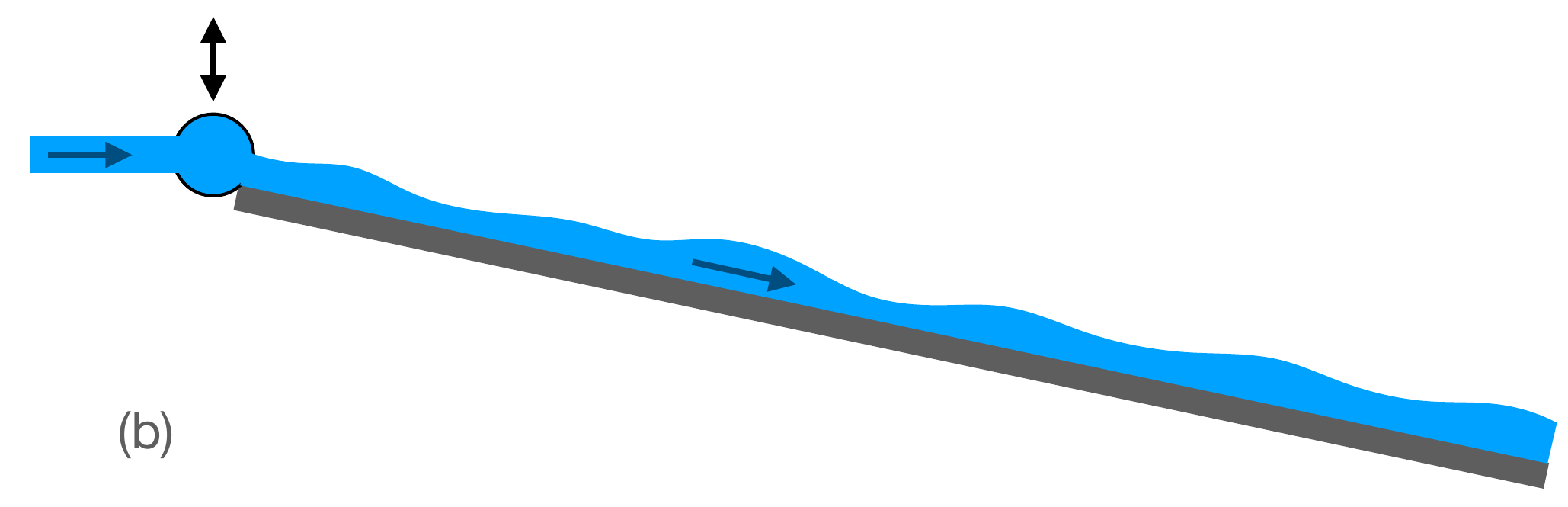}\
\caption{Panel (a): a sketch of the experiments on controlled transition in a boundary layer (adapted after Figs. 1,2 in \cite{kachanov1994physical}). Close to the inlet region, a ribbon band is embedded into the plate, allowing for the insertion of controlled perturbations in the flow. Panel (b): a sketch of the experiments on controlled transition for waves on a thin film flow over an inclined plate (adopted from Fig. 1 in \cite{liu1993measurements}). A perturbation to the flow at the inlet is performed by applying pressure 
variations to the entrance manifold.}
\label{fig:sketch}
\end{figure}

An important aspect of the space-time chaos and turbulence is the nature of the linear instability of a steady state (of a laminar flow). According to the classical stability theory~\cite{briggs1964electron,Lifshitz-Pitaevskii-81,Cross-Greenside-12} one distinguishes two types of instability in homogeneous nonequilibrium media:  absolute and convective.  In convectively unstable systems, a local in space perturbation grows but moves away from the place of its initiation, so that the perturbation in the region where it was imposed decays.  In absolutely unstable systems, a perturbation grows at the place where it was imposed. Convective instability appears in beam-plasma systems~\cite{briggs1964electron,manickam1975convective}, in electronics (e.g., in a traveling wave tube~\cite{Ginzburg-Pikovsky-Sergeev-89,kuznetsov2004wave}), in chemical reactions~\cite{Kuznetsov-Mosekilde-Dewel-Borckmans-97}. In hydrodynamics, convective instability appears in many open-flow systems, like wakes, capillary jets, and two-dimensional Poiseuille flow~\cite{huerre1990local}. Two examples we discuss in more detail below, are sketched in Fig.~\ref{fig:sketch}: boundary layer flow and waves on a liquid film over an inclined plate.

A common feature of convectively unstable systems is that one cannot directly apply notions of deterministic chaos, such as sensitive dependence on initial conditions, to them. Indeed, initial perturbations are ``blown away'' and are not relevant at large times when observing a fixed finite domain. Instead, one needs constantly acting perturbations close to the inlet zone to have a nontrivial dynamical regime~\cite{Deissler-85,Deissler-Kaneko-87,Pikovsky-88,Pikovsky-89b,Brand-Deissler-89,Deisler-Oron-Lee-91,lucke1993amplification,Rudzick-Pikovsky-96,Baia_etal25}. In experiments, two types of such perturbations have been explored. In many cases, chaos and turbulence develop from the natural, ambient noise. In other setups, one imposes an actuator that produces controlled perturbations. If these controlled perturbations are significantly larger than the level of natural noise, they dominate the close to the inlet domain of spatial growth and development. Typically, periodic in time perturbations are imposed to produce waves, properties of which are known from the theory, see~\cite{klebanoff1962three,kachanov1984resonant,kachanov1994physical} for such experiments with a boundary layer (Fig.~\ref{fig:sketch}(a)), and \cite{krantz1971stability,liu1993measurements,chang2002complex} for experiments with waves on liquid film (Fig.~\ref{fig:sketch}(b)). The resulting periodic structures can also be unstable, and natural noise leads to the loss of periodicity and the establishment of a turbulent state.

In a series of experiments~\cite{borodulin2011experimental,borodulin2013experimental,%
kachanov2013hypothesis,borodulin2014properties}, Borodulin, Kachanov, and co-workers used the same experimental setup for actuation of perturbation in a boundary layer Fig.~\ref{fig:sketch}(a) to impose  \textit{stochastic} macroscopic perturbations with controllable statistical properties. In these experiments, they discovered interesting features that enabled them to introduce the concepts of  ``deterministic'' and ``stochastic'' turbulence. These features appear to be common to systems with convective space-time chaos, and can be formulated independently of particular hydrodynamical aspects of the boundary layer transition.
 The following properties were emphasized by Borodulin, Kachanov, and co-workers:
\begin{enumerate}
\item Averaged
statistical characteristics (in particular, profiles of mean velocity and velocity fluctuations, and power spectra of velocity fluctuations) at sufficiently large distances from the inlet were the same for different realizations of the forcing signals.
\item A setup, where exactly the same random forcing signal was used in two different experiments. As a result, the measured turbulent fields in the two repetitions were very close to each other in a relatively large domain downstream of the forcing; this domain included both an initial region of perturbation growth and a part of the region with universal statistical properties, as described in feature 1. For example, in Fig. (3) of \cite{borodulin2014properties}, the authors present plots of two instantaneous fields of velocity perturbations measured in two experiments, one in the morning and one in the evening, with the same random force; these two fields nearly coincide. This regime is called deterministic turbulence.
\item The coincidence (measured by a cross-correlation of the signals) of two fields in two experiments, as described in feature 2, is not perfect. The uncorrelated component (which is unavoidable in experiments due to ambient noise) grows exponentially downstream (see Fig. (27) of \cite{borodulin2014properties}), and eventually, at large distances from the inlet, correlations between fields in two runs disappear. This regime is called stochastic turbulence.
\end{enumerate}

The goal of this paper is to give an interpretation of these features in terms of deterministic models of convective space-time chaos.   In particular, we will argue that the reproducibility (feature two above) directly corresponds to another phenomenon known in nonlinear dynamics -- reliability of randomly forced systems~\cite{Mainen-Sejnowski-95}. We emphasize here that our aim is not to model a boundary layer transition that is extremely challenging computationally, but rather to provide theoretical dynamical tools that can characterize the features of deterministic and stochastic turbulence as formulated above. These tools are not specific to hydrodynamics,  but can be generally implemented in other manifestations of convective space-time chaos, such as in plasma physics and electronics.

We start with formulating in Section
\ref{sec:smdst} two tractable models of convective space-time chaos, one as a system of ordinary differential equations, and another as a nonlinear partial differential equation. This partial differential equation describes waves on an inclined liquid film (Fig.~\ref{fig:sketch}(b)), which is a paradigmatic case of convective instability. In Section~\ref{sec:tcle} we present the main tool of the analysis, the convective Lyapunov exponent. Section \ref{sec:rf} describes the same setup of a macroscopic random driving at the inlet as one adopted in experiments~\cite{borodulin2011experimental,borodulin2013experimental,kachanov2013hypothesis,borodulin2014properties}. We describe how the features of deterministic and stochastic turbulence manifest themselves for the models of Section
\ref{sec:smdst}. In particular, we quantify the transition from deterministic to stochastic turbulence with the spatial Lyapunov exponent. Section \ref{sec:reg} is devoted to regular forcing. Especially for quasiperiodic driving (which is also feasible in experiments), one observes a nontrivial growth in signal complexity and an approach to effective ``randomness''. In Section~\ref{sec:std} we discuss spatio-temporal duality in convective space-time chaos, according to which one can exchange the meaning of space and time variables, to get better insight into the features of convective turbulence. We conclude with a discussion in Section~\ref{sec:concl}.

\section{Dynamical models for convective space-time chaos}
\label{sec:smdst}

\subsection{A partial derivative equation}
\label{sec:pde}
As a basic model for convective chaos continuous in space and time, we use the 
Kuramoto-Sivashinsky (KS)  equation. This equation was derived by Sivashinsky in 1977 for the instability of propagating flame fronts \cite{Sivashinsky-77} and by Kuramoto in 1976-78  for the phase dynamics beyond Benjamin-Fair-type instability in active nonlinear medium~\cite{kuramoto1976persistent,Kuramoto-78}.
Previously, this equation was derived in 1974 by Nepomnyashchy~\cite{Nepom-74} and Homsy~\cite{Homsy-74} for the nonlinear stage of instability of films flowing down an inclined plate, and in 1975 for plasma waves~\cite{laquey1975nonlinear}. The problem of waves on a flowing down film is mostly close to the boundary layer experiments~\cite{borodulin2011experimental,borodulin2013experimental,kachanov2013hypothesis,borodulin2014properties} (in fact, falling film can be considered as a boundary layer with an open surface) and will be adopted for the interpretation of the KS equation below. Correspondingly, we will refer to the basic equation as the Nepomnyashchy-Homsy-Kuramoto-Sivashinsky equation (NHKS).

Referring for the derivation of NHKS equation to~\cite{Nepom-74}, we write it here in dimensionless coordinates where all the coefficients except for one are set to one:
\begin{equation}
\frac{\partial u}{\partial t}+V\frac{\partial u}{\partial x}+u\frac{\partial u}{\partial x}+\frac{\partial^2 u}{\partial x^2}+\frac{\partial^4 u}{\partial x^4}=0\;.
\label{eq:ks1}
\end{equation}
Here, $u(x,t)$ represents the dimensionless deviation of the film width from its constant value; $x$ and $t$ are dimensionless space and time coordinates, respectively. Terms with higher derivatives in $x$ describe linear instability of perturbations. The lowest-order nonlinear term is responsible for the transformation of perturbations from the unstable to the stable part of the spectrum, which is sufficient for the saturation of the instability.

The only nontrivial parameter in this equation is the dimensionless velocity of linear waves $V$, which is represented via the geometry of the setup and the properties of the fluid as~\cite{Nepom-74}
\begin{equation}
\begin{gathered}
V=\frac{3 (5^{3/2}) \gamma^{1/2}G^{-1/3}\cos\alpha}{(2G\cos^2\alpha-\sin\alpha)^{3/2}},\quad
G=\frac{ga_0^2}{\nu^2},\\
 \gamma=\frac{\sigma}{\nu^{4/3}g^{1/3}\rho}\;.
\end{gathered}
\label{eq:v}
\end{equation}
Here $\alpha$ is the angle of the plate to the vertical direction, $G$ and $\gamma$ are the Galilei and the Kapitza numbers; $g$ is the gravity acceleration, $\nu$ is the kinematic viscosity, $\rho$ is the density of fluid, $\sigma$ is the surface tension coefficient, 
$a_0$ is the depth of the unperturbed film. Equation \eqref{eq:ks1} is written for the case of instability, the threshold of which is defined by the condition $G>G_*= \frac{5\sin\alpha}{2\cos^2\alpha}$. The NHKS equation is valid close to the instability threshold; thus, according to \eqref{eq:v}, velocity $V$ is large. However, for the sake of easy computations, below we consider not so large values of $V$; qualitatively, all the effects remain the same.

In numerical simulations we use a discretization of \eqref{eq:ks1} with spatial step $h$:
\begin{equation}
\begin{gathered}
\frac{d u_k}{dt}+(6h)^{-1}(u_{k-1}+u_k+u_{k+1}+3V)(u_{k+1}-u_{k-1})+\\
h^{-2}(u_{k-1}-2u_k+u_{k+1})+\\
h^{-4}(u_{k-2}-4u_{k-1}+6u_k-4u_{k+1}+u_{k+2})=0\;.
\end{gathered}
\label{eq:ksdiscr}
\end{equation}
Below, we solve this ODE system using the Runge-Kutta 4th-order method with time step $2\cdot10^{-4}$ and $h=0.2$. This spatial step is small enough to ensure spatial smoothness of the solution.

Equation \eqref{eq:ks1} should be supplemented with boundary and initial conditions. It is natural to set the inlet of the film to $x=0$ and consider the downstream domain $0\leq x<\infty$. Then, equation \eqref{eq:ks1} should be supplemented by the initial condition $u(0\leq x<\infty,0)$ and the boundary conditions at $x=0$. If one considers a finite domain $0\leq x\leq L$, then also boundary conditions at $x=L$ are needed. Physically, these conditions should describe external forcing at the inlet $x=0$ and no-reflection at $x=L$. Because the NHKS equation contains higher spatial derivatives, the boundary conditions should include some conditions on derivatives. For the discrete version \eqref{eq:ksdiscr}, to solve the equations for $1\leq k\leq N$, we have to define $u_0,u_{-1},u_{N+1},u_{N+2}$. At the inlet we adopted $u_0=f(t)$, $u_{-1}=u_1$, where $f(t)$ is the external forcing of the system.
At the outlet we adopted $u_{N+1}=2u_N-u_{N-1}$ and $u_{N+2}=2u_{N}-u_{N-2}$. This corresponds to the vanishing of the second and fourth discrete derivatives. Although this does not ensure a perfect non-reflection, distortions are restricted to a relatively narrow boundary domain.

\begin{figure}[!htb]
\centering
\includegraphics[width=\columnwidth]{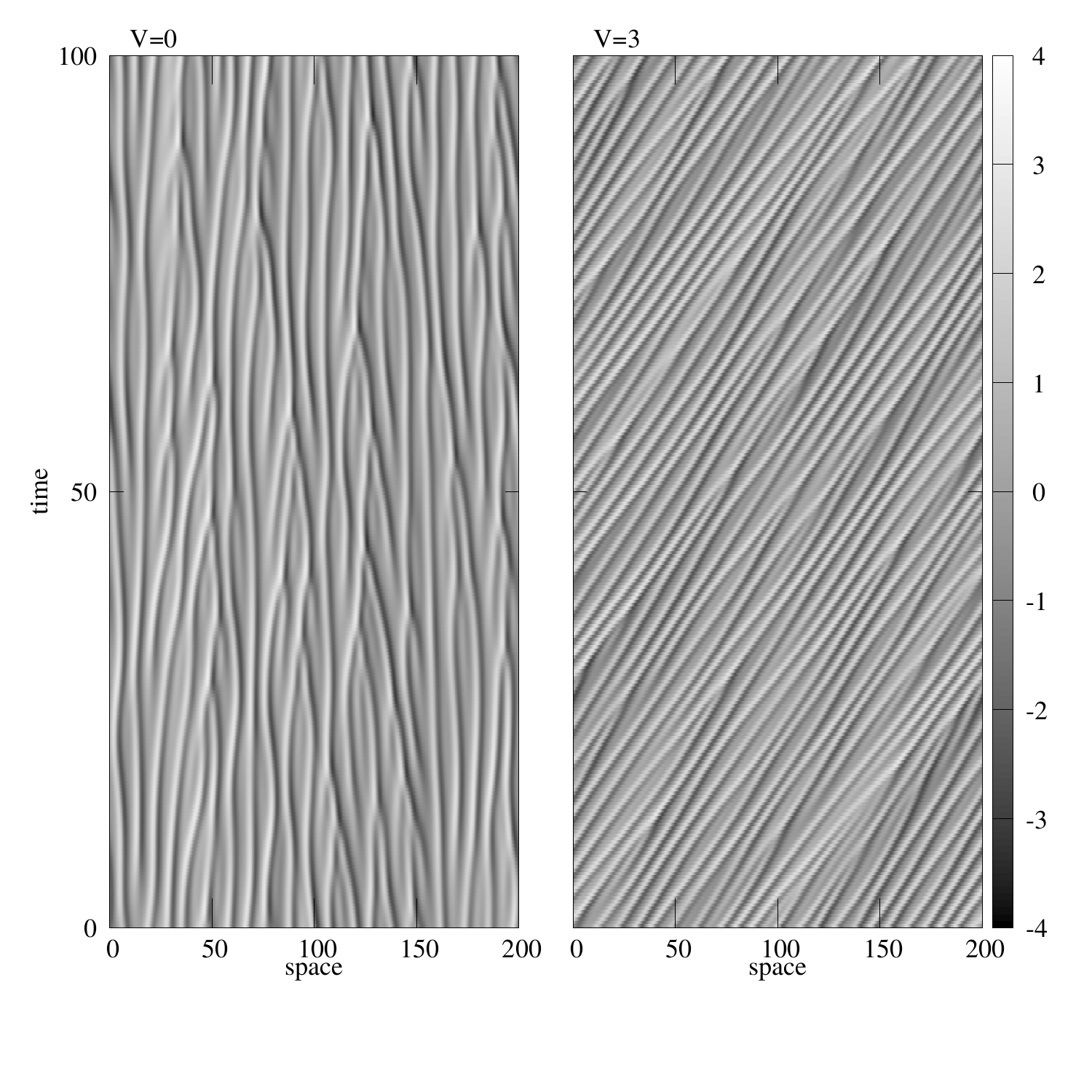}
\caption{Space-time diagrams of $u(x,t)$ for different $V$ and periodic boundary conditions, for $V=0$ (Eq.~\eqref{eq:ks2}) and $V=3$ (Eq.~\eqref{eq:ks1}).}
\label{fig:ksfield}
\end{figure}

To consider properties of the wave dynamics in NHKS equation in a free regime away from the boundaries, one often imposes periodic boundary conditions in a finite domain $u(x,t)=u(x+L,t)$. Furthermore, in such a regime, one can eliminate the parameter $V$ by performing a transformation to a reference system moving with a linear velocity $V$. Then one obtains a ``standard'' Kuramoto-Sivashinsky equation
\begin{equation}
\frac{\partial u}{\partial t}+u\frac{\partial u}{\partial x}+\frac{\partial^2 u}{\partial x^2}+\frac{\partial^4 u}{\partial x^4}=0\;,\quad u(x,t)=u(x+L,t)\;,
\label{eq:ks2}
\end{equation}
solutions of which depend only on the dimensionless domain size $L$. For large enough $L$, these solutions are chaotic, and Eq.~\eqref{eq:ks2} represents one of the paradigmatic models for space-time chaos~\cite{Manneville-88,bohr1998dynamical,lan2008unstable}. We illustrate chaotic states in equations \eqref{eq:ks1} and \eqref{eq:ks2} (in both cases periodic boundary conditions) in Fig.~\ref{fig:ksfield}.

We conclude this section by mentioning that Eq.~\eqref{eq:ks1} is valid also for phase turbulence in the complex Ginzburg-Landau equation. The latter equation describes the evolution of a complex amplitude $A(x,t)$ of an unstable wave close to the instability threshold, it reads (cf. eq. (10.63) in~\cite{Cross-Greenside-12})
\begin{equation}
(\partial_t+s\partial_x)A=A+(1+c_1)\partial_x^2 A-(1-ic_3)|A|^2A\;.
\label{eq:gl1}
\end{equation}
A spatially homogeneous solution with amplitude $|A|=1$ is stable if $1-c_1c_3>0$ (Newell's criterion) but becomes unstable if $1-c_1c_3<0$. Close to this threshold, the real amplitude $|A|$ remains close to one, but the phase $\theta(x,t)=\text{arg}(A)$ is modulated. To describe these modulations, one reduces to a single equation for the dynamical variable $\theta$, assuming that the amplitude is slaved by the variations of the phase~\cite{Kuramoto-78}. Such a reduction is, e.g., performed in Refs.~\cite{baalen2004phase,vercesi2024scaling}; it yields
\begin{equation}
\partial_t\theta+s\partial_x\theta-c_3-(1-c_1c_3)\partial_x^2\theta+\frac{c_1^2}{2}\partial_x^4\theta+(c_1+c_3)(\partial_x\theta)^2=0\;.
\label{eq:gl2}
\end{equation}
By setting $u(x,t)=\partial_x\theta(x,t)$, after a proper rescaling of the variables, this equation reduces to Eq.~\eqref{eq:ks1}.
\subsection{A chain of nonlinear amplifiers}
\label{sec:ode}
The PDE model of Section~\ref{sec:pde} is continuous in space and time. One can reduce complexity and numerical efforts by considering a discrete space and continuous time, and by formulating a set of ordinary differential equations (ODEs) that exhibit features similar to those of PDE \eqref{eq:ks1}. We consider a ``chain of nonlinear amplifiers'' \cite{Pikovsky-88,Pikovsky-89b}
\begin{equation}
\begin{aligned}
\frac{du_n}{dt}&=-u_n+F(u_{n-1}),\quad n=2,3,\ldots,N\;,\\
\frac{du_1}{dt}&=-u_1+f(t)\;.
\end{aligned}
\label{eq:ca}
\end{equation}
Here, each unit is a low-pass linear filter, driven by the transformed field of the left neighbor. As a transformation function, we use the logistic map $F(u)=1-2u^2$. The first unit is driven by a given external force $f(t)$. The asymmetric in space coupling describes the propagation of perturbations from small to larger values of $n$. Equations \eqref{eq:ca} should be supplemented with initial conditions $u_n(0)$. If these values are close to zero, and the force is bounded $|f(t)|<1$, then the solutions at all $n$ remain in this interval $|u_n|\leq 1$. 

Similar to the PDE of Section~\ref{sec:pde}, the chain \eqref{eq:ca} can be considered without forcing but with periodic boundary conditions $u_0(t)=u_N(t)$. Then, the only parameter is the system size $N$. For $N\geq 6$, typically, space-time chaos is observed.

\subsection{A coupled map lattice}
\label{sec:cml}
Here we describe a maximally simplified version of convective chaos, where both space (variable $n$) and time (variable $j$) are discrete. This is a coupled map lattice (CML) model with a unidirectional coupling
\begin{equation}
\begin{gathered}
u(j+1,n)=(1-\epsilon)F(u(j,n))+\epsilon F(u(j,n-1)),\\ n=1,2,\ldots;\quad u(j,0)=f(j)\;.
\end{gathered}
\label{eq:cml}
\end{equation}
Here, the coupling parameter $\epsilon\in [0,1]$ plays the role of the velocity parameter $V$ in the PDE model \eqref{eq:ks1}; large $\epsilon$ corresponds to a strong advection of perturbations compared to the local dynamics. For the nonlinear map $F(u)$, one typically takes the logistic map. The term $f(j)$ describes discrete-time forcing at the inlet.  Model~\eqref{eq:cml} has been explored in \cite{Rudzick-Pikovsky-96,Baia_etal25}. The main advantage of the discrete model \eqref{eq:cml} compared to continuous models \eqref{eq:ca} and \eqref{eq:ks1} is the relative simplicity of numerical simulations; however, this advantage is no longer essential nowadays. Thus, we will not follow the CML model \eqref{eq:cml} below in this paper, concentrating instead on realistic and experimentally relevant models \eqref{eq:ks1} and \eqref{eq:ca}.

\section{Temporal and convective Lyapunov exponents}
\label{sec:tcle}

Lyapunov exponents (LEs) are the main theoretical and numerical tool for characterization of chaos via quantification of sensitive dependence of the dynamics on initial conditions~\cite{Pikovsky-Politi-16}. In spatio-temporal dynamics, in the case of a spatially infinite medium, the LEs are not well-defined, because different norms used to characterize the size of perturbations are not equivalent. One way to circumvent this is to consider large finite systems with periodic boundary conditions. Then, the usual largest LE can be determined, which converges to a definite value as the length of the system goes to infinity~\cite{Pikovsky-Politi-98,Pikovsky-Politi-16}. To distinguish this usual LE from other types, we will call it \textit{temporal} LE. 

More detailed information on how linear perturbations $U(x,t)$ on top of the chaotic dynamics
$u(x,t)$ grow and propagate is delivered by the convective (velocity-dependent) LE~\cite{Deissler-Kaneko-87, Pikovsky-Politi-16}. To define it, one considers an infinite system, makes a local perturbation at site $x=0,t=0$, and looks at the exponential growth of this perturbation along the ray $x=vt$:
\begin{equation}
\lambda(v)=\lim_{T\to\infty} \frac{1}{T}\ln \frac{||U(vT,T)||}{||U(0,0)||}\;.
\label{eq:convle}
\end{equation}
Practically, finite time intervals $T$ and thus also finite domains are considered, so that there is no issue of norms' non-equivalence. Two values of the convective LE are of special importance. The maximal value $\lambda_{max}$ is the same as the temporal LE in a large spatial domain with periodic boundary conditions. The value at zero velocity $\lambda(0)$ is the temporal LE in a finite spatial domain with non-reflecting (open) boundary conditions.

The case where $\lambda_{max}>0$ and $\lambda(0)<0$ is that of convective space-time chaos. Typically, in the case of convective space-time chaos, the trivial steady state is also linearly convectively unstable. Thus, for observation of space-time chaos, one has to apply perturbations at the inlet boundary (which boundary is the inlet one, is determined by the sign of velocity $v_{max}$ at which $\lambda_{max}=\lambda(v_{max})$).

We now analyze the convective LE for models \eqref{eq:ks1} and \eqref{eq:ca}.
The best way to calculate the convective LE is not to apply Eq.~\eqref{eq:convle}, but instead to calculate the usual temporal largest LE for exponential in space perturbations, and then to apply the Legendre transform~\cite{lepri1996chronotopic,Pikovsky-Politi-16}. One considers a large system with periodic boundary conditions, and follows linear perturbations in the form $e^{\mu x} U(x,t)$ with periodic boundary conditions for $U$. The exponential growth rate of the norm of $U$ yields the $\mu$-dependent LE $\mathcal{L}(\mu)$ (for $\mu=0$ the usual temporal LE is obtained), which we will call chronotopic exponent following terminology of \cite{lepri1996chronotopic,Pikovsky-Politi-16}. Then, the convective LE is obtained via the Legendre transform~\cite{lepri1996chronotopic,Pikovsky-Politi-16}
\begin{equation}
\lambda(v)=\mathcal{L}(\mu)+\mu v,\qquad v=-\frac{d\mathcal{L(\mu)}}{d\mu}\;.
\label{eq:expconv}
\end{equation}
Because the Legendre transform is invertible, one can also express the chronotopic exponent via the convective one:
\begin{equation}
\mathcal{L}(\mu)=\lambda(v)-\mu  v,\qquad \mu=\frac{d\lambda(v)}{dv}\;,
\label{eq:expconvinv}
\end{equation}
we will use this expression in the discussion of the spatial LE below in Section~\ref{sec:sple}.

In terms of the chronotopic exponent $\mathcal{L}(\mu)$, the velocity-dependent exponent at zero velocity is $\lambda(0)=\min_{\mu} \mathcal{L}(\mu)$, while the maximal velocity-dependent exponent is, according to \eqref{eq:expconvinv}, $\lambda_{max}=\mathcal{L}(0)$. Thus, the conditions of convective space-time chaos $\lambda_{max}>0,\;\lambda(0)<0$ in terms of the chronotopic exponent can be formulated as the property that $ \mathcal{L}(\mu)$ changes sign at some $\mu$. 

Mostly simply this approach leads to the convective LE for the chain of amplifiers \eqref{eq:ca}. Indeed, the linearized equations on top of spatio-temporal state $u_n(t)$
read
\begin{equation}
\frac{dU_n}{dt}=-U_n+F'(u_{n-1}(t))U_{n-1}\;.
\label{eq:linca}
\end{equation}
Substituting here $U_n(t)=e^{\mu n} e^{-t} z_n(t)$ leads to a linear system
\begin{equation}
\frac{dz_n}{dt}=e^{-\mu}F'(u_{n-1}(t))z_{n-1}\;,
\label{eq:z}
\end{equation}
in which the time can be rescaled $t\to e^{-\mu} t'$. Thus, the $\mu$-dependent chronotopic LE is
\begin{equation}
\mathcal{L}(\mu)=-1+\lambda_0e^{-\mu}\;,
\label{eq:mudle}
\end{equation}
where $\lambda_0$ is the LE of \eqref{eq:z} for $\mu=0$. Performing the Legendre transform, we obtain the convective exponent
\begin{equation}
\lambda(v)=-1+v-v\ln\frac{v}{\lambda_0}\;.
\label{eq:convleca}
\end{equation}

\begin{figure}[!htb]
\centering
\includegraphics[width=\columnwidth]{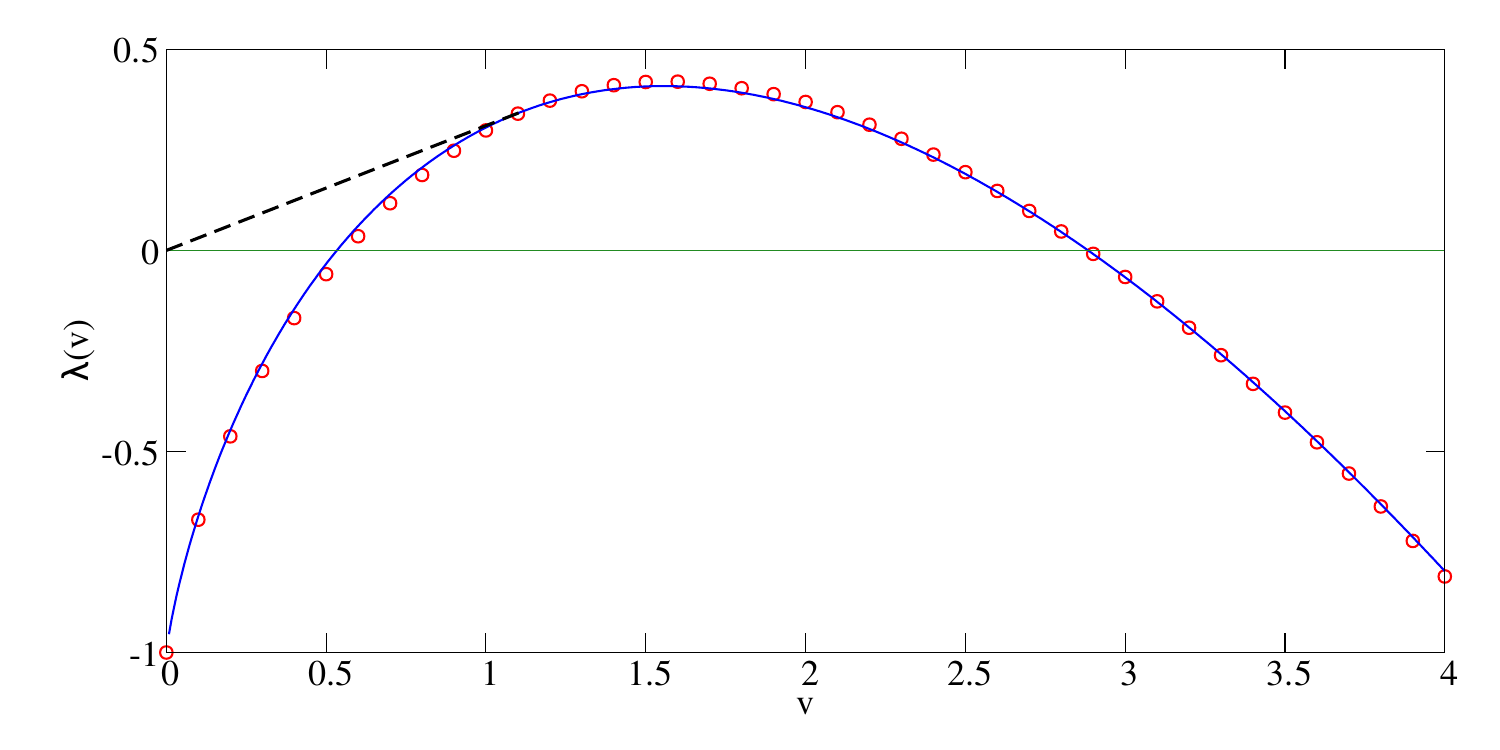}
\caption{Convective LE for model \eqref{eq:ca}. Blue line is fit $\lambda(v)=-1+1.305 v -0.905 v\ln v$. The dashed black line shows the construction of the spatial LE according to Eq.~\eqref{eq:forsle}.}
\label{fig:convleca}
\end{figure}

Numerics of the convective exponent  (Fig.~\ref{fig:convleca}) yields the best fit with slightly distorted expression \eqref{eq:convleca}, as indicated in the caption.  The usual LE here is $\lambda(0)=-1$. This result is evident because the linear matrix in \eqref{eq:linca} is a lower triangular matrix with all diagonal elements equal to $-1$. The largest LE $\lambda_{max}\approx 0.41$ is achieved at $v\approx 1.55$ and is positive. Thus, the system belongs to the class of convective space-time chaos.

\begin{figure}[!htb]
\centering
\includegraphics[width=\columnwidth]{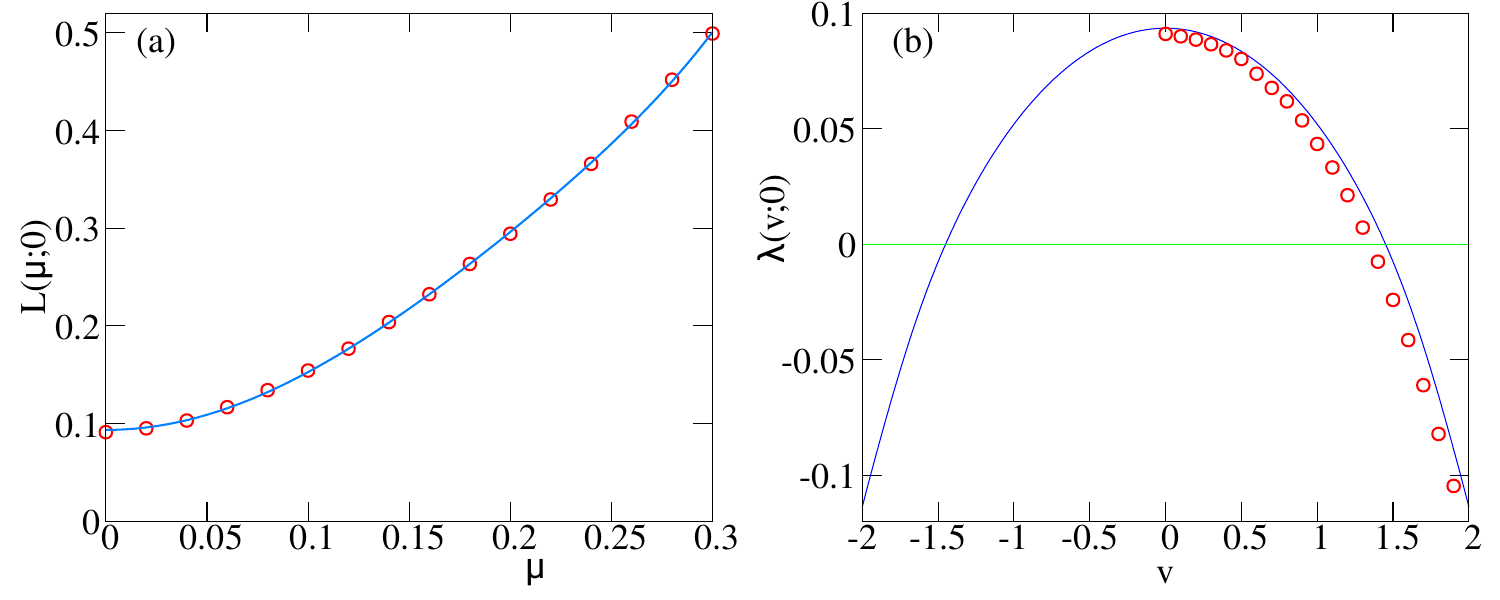}
\caption{Panel (a): The chronotopic LE $\mathcal{L}(\mu;0)$. Markers: numerical results; solid line: polynomial fit $\mathcal{L}(\mu;0)=a-bx^2-cx^4-dx^6$ with $a               = 0.0935; b=6.29; c=-39.4; d=219.7$. Panel (b): The convective Lyapunov exponent $\lambda(v;0)$. Markers: direct numerical calculations according to the definition \eqref{eq:convle}, solid blue line - Legendre transform of the polynomial fit of $\mathcal{L}(\mu;0)$. The maximal speed of growing perturbations is $v_0\approx 1.4$.}
\label{fig:convleks}
\end{figure}

For the NHKS equation \eqref{eq:ks1}, the chronotopic and the convective exponents depend on parameter $V$: $\mathcal{L}(\mu;V)$, $\lambda(v;V)$. These dependences are, however, trivial as we demonstrate below. Indeed, in the linearized NHKS equation \eqref{eq:ks1} the term with $V$ leads to a damping with factor $-\mu V$, thus $\mathcal{L}(\mu;V)=\mathcal{L}(\mu;0)-\mu V$. Then, according to application of the Legendre transform~\eqref{eq:expconv}, the convective LE has the form $\lambda(v;V)=\lambda(v-V;0)$. This reflects the fact that the NHKS equation~\eqref{eq:ks1} and the standard KS equation \eqref{eq:ks2} are related by a transformation to the reference frame moving with velocity $V$.

Calculations of the chronotopic LE for the KS equation are presented in Fig.~\ref{fig:convleks}. The maximal speed of growing perturbations is $v_0\approx 1.4$. Thus, in the NHKS for $V>v_0$, one observes convective space-time chaos. We mention here that the maximal propagation speed of growing linear perturbations on top of the state $u=0$ in the KS equation is~\cite{conrado1994singular,conrado1995growth} $v_{lin}\approx 1.6$, i.e., slightly larger than $v_0$.

We conclude this section by summarizing that both the discrete model \eqref{eq:ca} and the PDE model \eqref{eq:ks1} for $V>v_0$ possess convective LE which has a positive maximum at some positive velocity, and is negative at zero velocity. This means that a nontrivial regime is possible only if a driving force is applied at the inlet. Below we consider cases of random and regular forces in Sections~\ref{sec:rf} and \ref{sec:reg}.

\section{Random input: noise-sustained turbulence, reliability, and spatial Lyapunov exponent}
\label{sec:rf}
In this and in the next Section, we consider finite systems \eqref{eq:ks1} and \eqref{eq:ca}
in the regime where the usual temporal LE is negative. This means that to have a nontrivial state, a non-constant driving force $f(t)$ has to be applied. In this Section, we assume this force to be a stationary random process. 

\subsection{Noise-sustained turbulence}
\label{sec:nst}
Due to the amplification and nonlinear transformation of perturbations from the boundary toward the bulk of the medium, one expects spatial evolution of the statistical properties of the process and eventual establishment of a statistically stationary in space regime with the same properties as those in a large system with periodic boundary conditions.  This phenomenon of ``noise-sustained turbulence'' has been demonstrated in \cite{Deissler-85} for the Ginzburg-Landau equation and in \cite{Brand-Deissler-89} for a variant of Eq.~\eqref{eq:ks1} that includes a linear damping term.

\begin{figure}[!htb]
\centering
\includegraphics[width=\columnwidth]{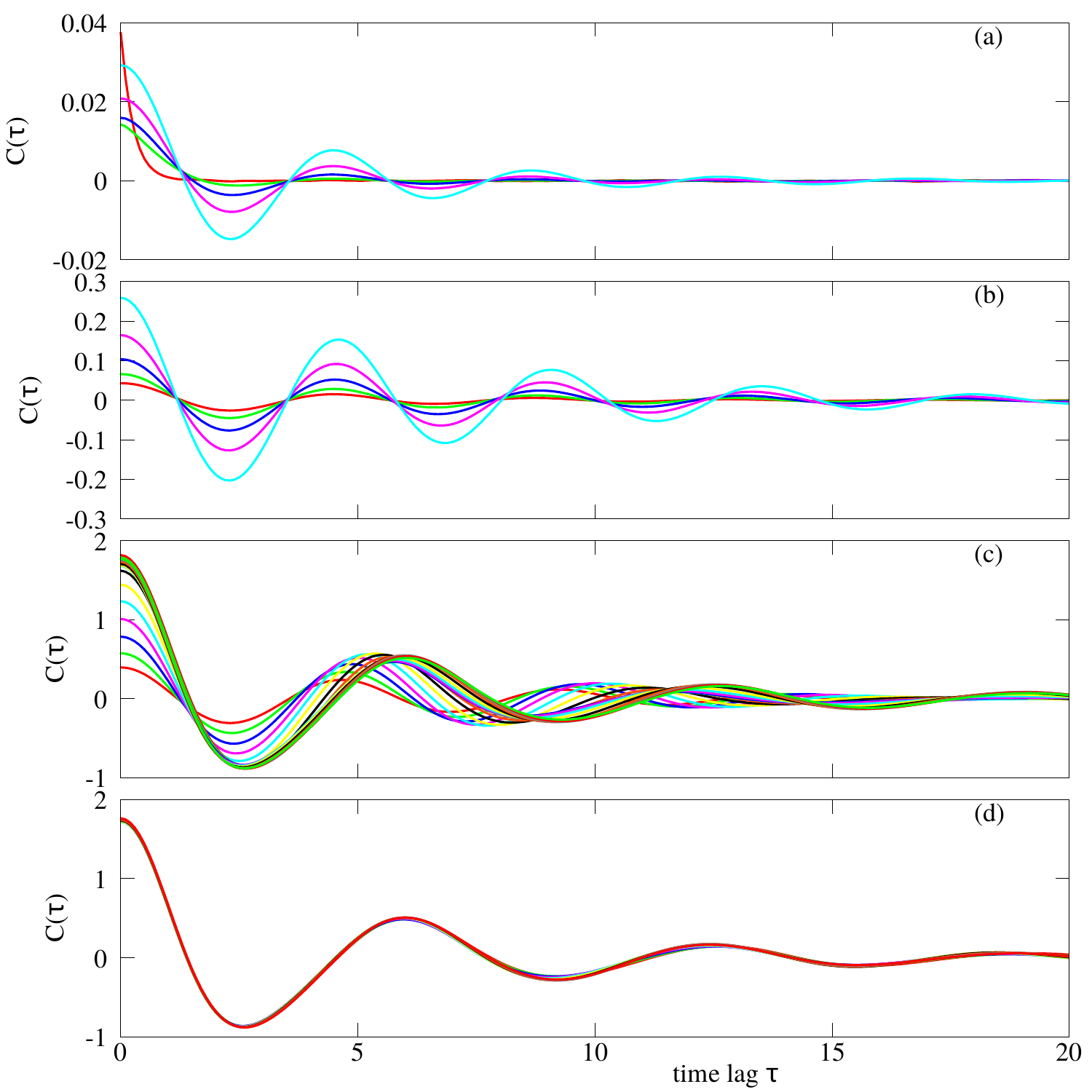}
\caption{The autocorrelation functions for different spatial positions for an OU-process driving force in the NHKS model \eqref{eq:ks1} with $V=2$.   Panel (a): $x=0,2,4,6,8$, Panel (b): $x=10,12,14,16,18$, Panel (c): $x=20,\ldots,58$; Panel (d): $x=60,\ldots,78$. }
\label{fig:nstks}
\end{figure}

We illustrate noise-sustained turbulence in the NHKS equation \eqref{eq:ks1} in Fig.~\ref{fig:nstks}. As a noisy force, we use an Ornstein-Uhlenbeck stationary Gaussian process with zero mean and exponential autocorrelation function $C_f(\tau)=\langle f(t)f(t+\tau)\rangle=0.04\exp[-4\tau]$. For characterizing the statistical characteristics of the field $u(x,t)$ at different distances from the inlet, we also used an autocorrelation function 
$C(\tau;x)=\langle (u(x,t)-\langle u\rangle )(u(x,t+\tau)-\langle u\rangle )\rangle$. The spatial evolution of the random field has two features: (i) the variance grows (because the driving force has a small amplitude in this example) and saturates at large distances from the inlet; (ii) the shape of the autocorrelation function changes from monotonous exponential to oscillatory exponential. One can see from the bottom panel of Fig.~\ref{fig:nstks} that the autocorrelation functions for all $x>60$ practically coincide, indicating the onset of a spatially statistically homogeneous space-time chaos.

A similar evolution of a random driving force along the chain of nonlinear amplifiers is observed in the discrete model \eqref{eq:ca}.

\subsection{Synchronization by common noise and reliability}
\label{sec:rel}

As discussed in Section~\ref{sec:tcle}, in the regimes of noise-sustained turbulence (Section~\ref{sec:nst}), the usual temporal LE is negative. This allows for application of the concept of synchronization by common noise~\cite{Pikovsky-84,Pikovsky-84a,Pikovsky-Rosenblum-Kurths-01,Goldobin-Pikovsky-04,Goldobin-Pikovsky-05a} to these regimes. This effect appears if the same noise drives two identical systems. Then, due to symmetry, a regime exists where the states of two systems coincide identically. The usual temporal LE governs the stability of this synchronous state. If it is negative, the synchronous state is stable, and typically establishes after some transient if the initial conditions of the two driven systems are different. 

Application of this concept to models \eqref{eq:ks1} and \eqref{eq:ca} implies that if two identical replicas of NHKS equation \eqref{eq:ks1} or of the chain of amplifiers \eqref{eq:ca} are driven by the same noisy forcing, then the states of these replicas will be identical.

The concept of synchronization by common noise is deeply related to the concept of reliability~\cite{Mainen-Sejnowski-95}. Here, instead of making a replica of a driven nonlinear system, one applies the same recorded noise signal several times to one system (in particular, as discussed in~\cite{Mainen-Sejnowski-95}, to one neuron). Then, if the LE is negative, the same response (after some initial transient depending on uncontrolled initial conditions) will be observed in different runs. Exactly this type of experiment has been performed at a random driving of the boundary layer in Refs.~\cite {borodulin2011experimental,borodulin2013experimental}.  

\begin{figure}[!htb]
\centering
\includegraphics[width=\columnwidth]{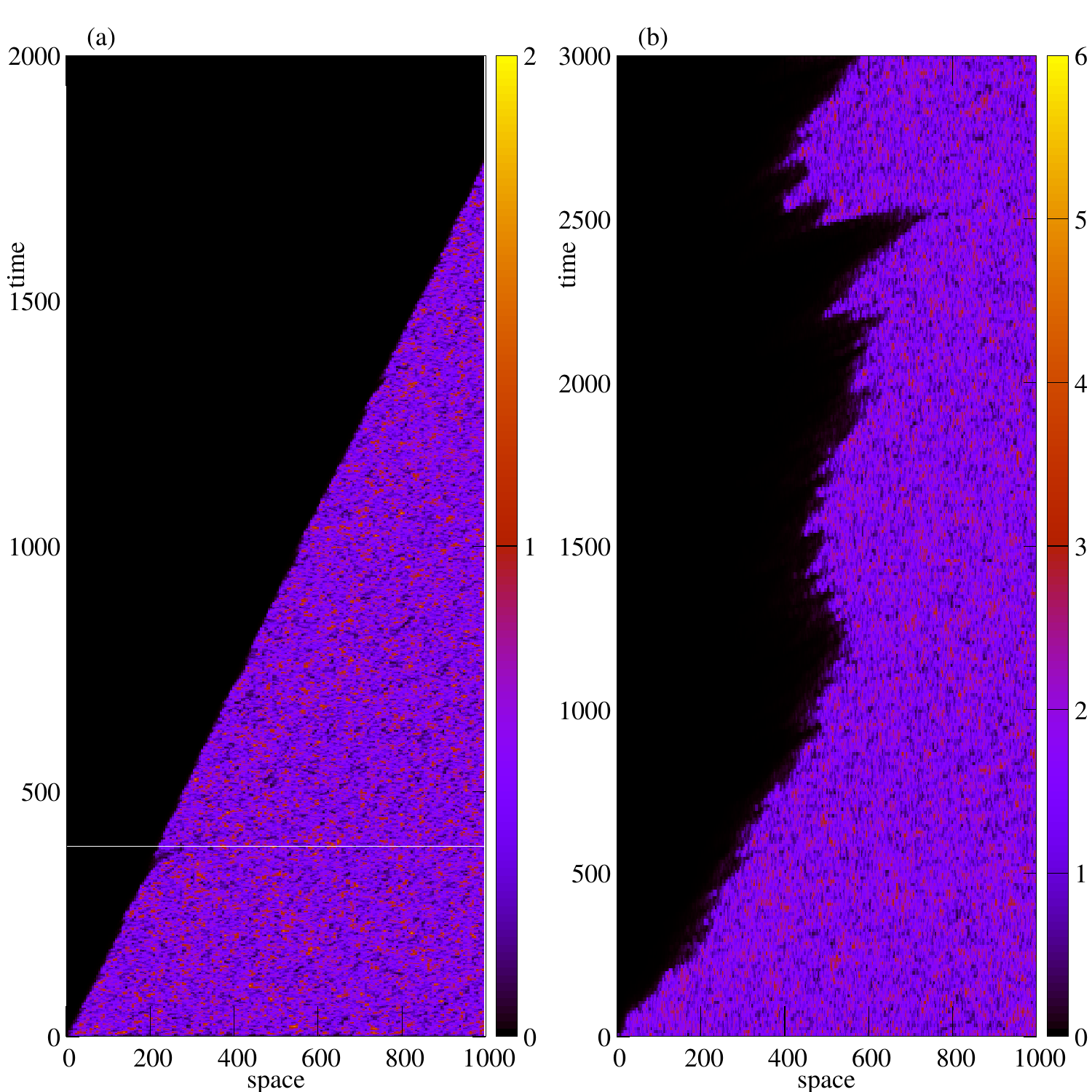}
\caption{Spatio-temporal plots of the mismatch between two replicas of the noisy driven systems with convective space-time chaos. Panel (a): for the chain of ODEs~\eqref{eq:ca}, a perfect synchrony is observed for all distances from the inlet. Panel (b): for the NHKS equation \eqref{eq:ks1} at large distances $x\gtrsim 500$ synchrony is broken.}
\label{fig:syn}
\end{figure}

We illustrate synchronization by common noise in Fig.~\ref{fig:syn}. For the ODE model \eqref{eq:ca} we show in panel (a) the space-time plot of the difference $|u_n(t)-v_n(t)|$ of the solution of \eqref{eq:ca} and of the solution of the replica 
\[
\frac{dv_n}{dt}=-v_n+F(v_{n-1}),\qquad
\frac{dv_1}{dt}=-v_1+f(t)\;.
\]
Both systems are driven by the same Ornstein-Uhlenbeck noise $f(t)$ with mean value $0.6$ and the exponential autocorrelation function $C(\tau)=0.01\exp[-4\tau]$. Initial states $v_n(0)$ and $u_n(0)$ are random and uncorrelated. One can see how the synchronous state $v_n(t)=u_n(t)$ (black color) establishes along the chain, via a propagation toward larger values of the space variable $n$.

A similar simulation for the NHKS model \eqref{eq:ks1} and its replica (in terms of variable $v(x,t)$) shows that synchrony establishes only in a domain close to the inlet $x\lesssim 500$, while for larger $x$ the fields of the system $u(x,t)$ and of its replica $v(x,t)$ remain different. We will explain this feature in the next Section~\ref{sec:sple}.

\subsection{Stability of synchronous state and spatial Lyapunov exponent}
\label{sec:sple}

The stability of synchrony in the presence of common noise is based on the negative LE. This means that if the symmetry between the system and its replica is slightly violated (e.g., due to small differences in the applied forces, or due to a small mismatch of the parameters of the system and the replica), then the states of the system and the replica will be slightly different~\cite{Goldobin-Pikovsky-05b}. 

However, in the considered case of convective space-time dynamics, there is an additional factor that influences stability. As has been discussed in Section~\ref{sec:tcle}, if there is a small local in space and time perturbation close to the inlet, then it will grow and propagate, and only after this growing pulse reaches the right boundary, the perturbation disappears. Suppose now that there are constant in time (in a statistical sense) perturbations at the inlet. They will be amplified along the medium, and we can characterize this amplification by the \textit{spatial} LE~\cite{Pikovsky-Politi-16,Baia_etal25} defined as
\begin{equation}
\Lambda=\lim_{X\to\infty}\frac{1}{X}\ln \frac{||U(X,t)||_t}{||U(0,t)||_t}\;.
\label{eq:sple}
\end{equation}
Here $||\cdot||_t$ denotes a norm of a function of time $t$, defined, e.g., over a large finite time interval.

\begin{figure}[!htb]
\centering
\includegraphics[width=\columnwidth]{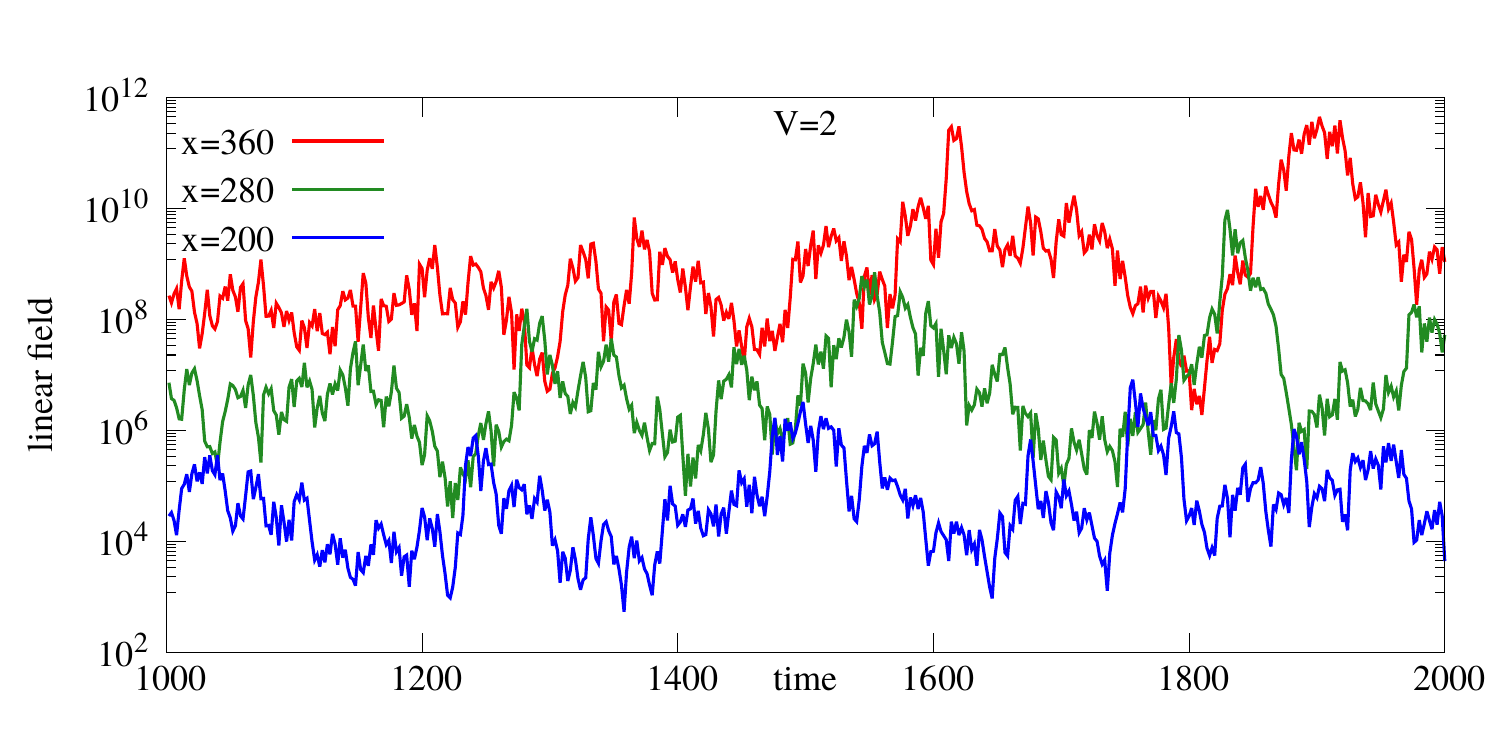}
\caption{Time dependencies of $\tilde U(x,t)$ at different $x$ for some values of velocity paramerter $V$. Parameter of temporal coarse-graining $\tau=2$.}
\label{fig:spfield}
\end{figure}

Here, we illustrate how to calculate this exponent in practice, for the NHKS model~\eqref{eq:ks1}. Having a space-time chaotic solution $u(x,t)$ (which we assume to be in a statistically stationary, typical case of space-time chaos; see however, a discussion at the end of this section), the linearized problem is formulated as a PDE
 \begin{equation}
\partial_t U+V\partial_x U+U\partial_x u(x,t)+u(x,t)\partial_x U+\partial_{xx}U+\partial_{xxxx}U=0 \;.
 \end{equation}
This equation is subject to the initial $U(x,0)$ and to boundary conditions. We use the same boundary conditions as for the NHKS equation (cf. discussion after Eq.~\eqref{eq:ksdiscr}), with some arbitrary stationary in time function $f(t)$ (e.g., one can just take the same noisy driving as used in Fig.~\ref{fig:syn}(b)). Then, at different positions $x$ at large enough (so that the initial conditions move away) times, one observes statistically stationary in $t$ fields; we illustrate this in Fig.~\ref{fig:spfield}. Because only the norm is important, we coarse-grain the fields $U(x,t)$ by calculating 
\[
\tilde U(x,t)=\left(\frac{1}{\tau}\int_{t-\tau/2}^{t+\tau/2} U^2(x,t')dt'\right)^{1/2}\;,
\]
this allows us to avoid singularities in the logarithmic representation.

\begin{figure}[!htb]
\centering
\includegraphics[width=\columnwidth]{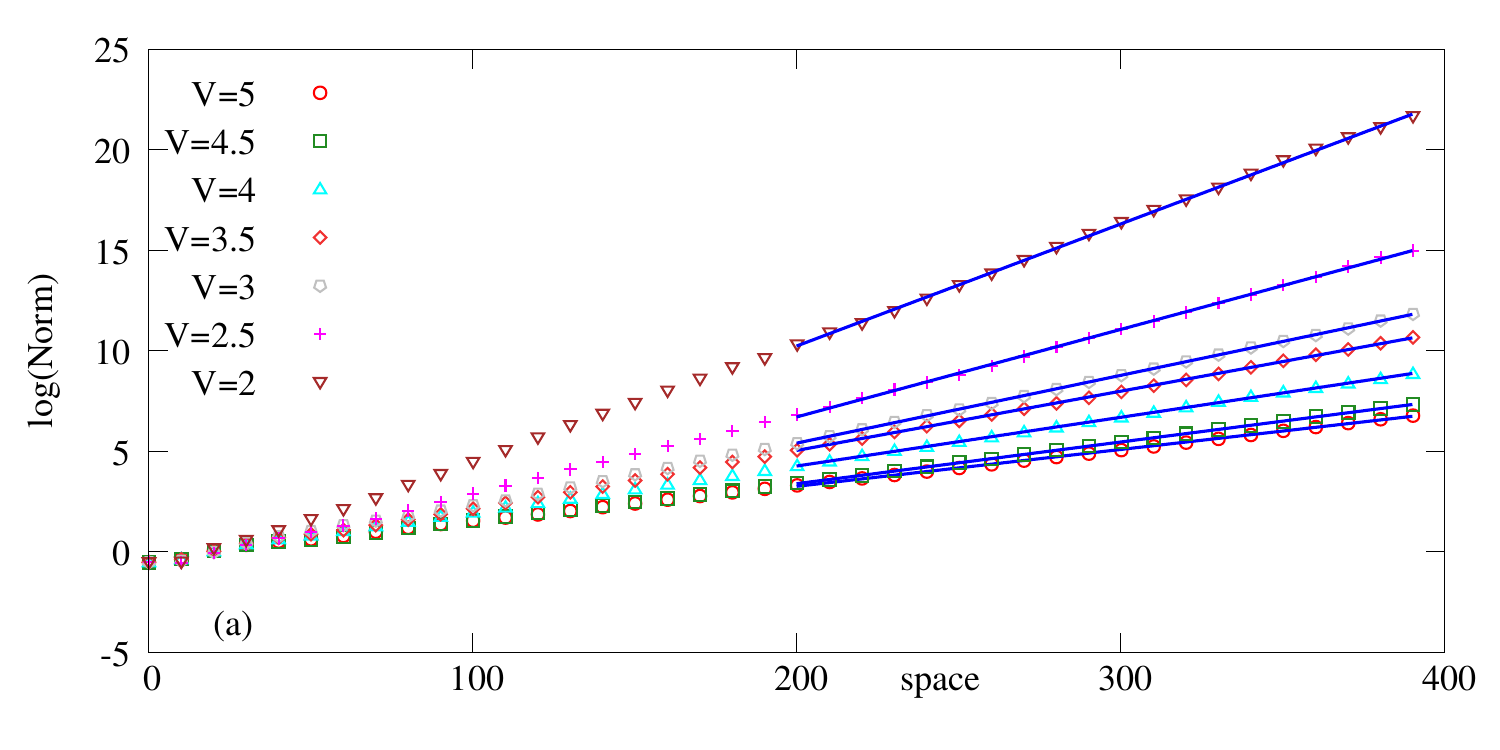}\\
\includegraphics[width=\columnwidth]{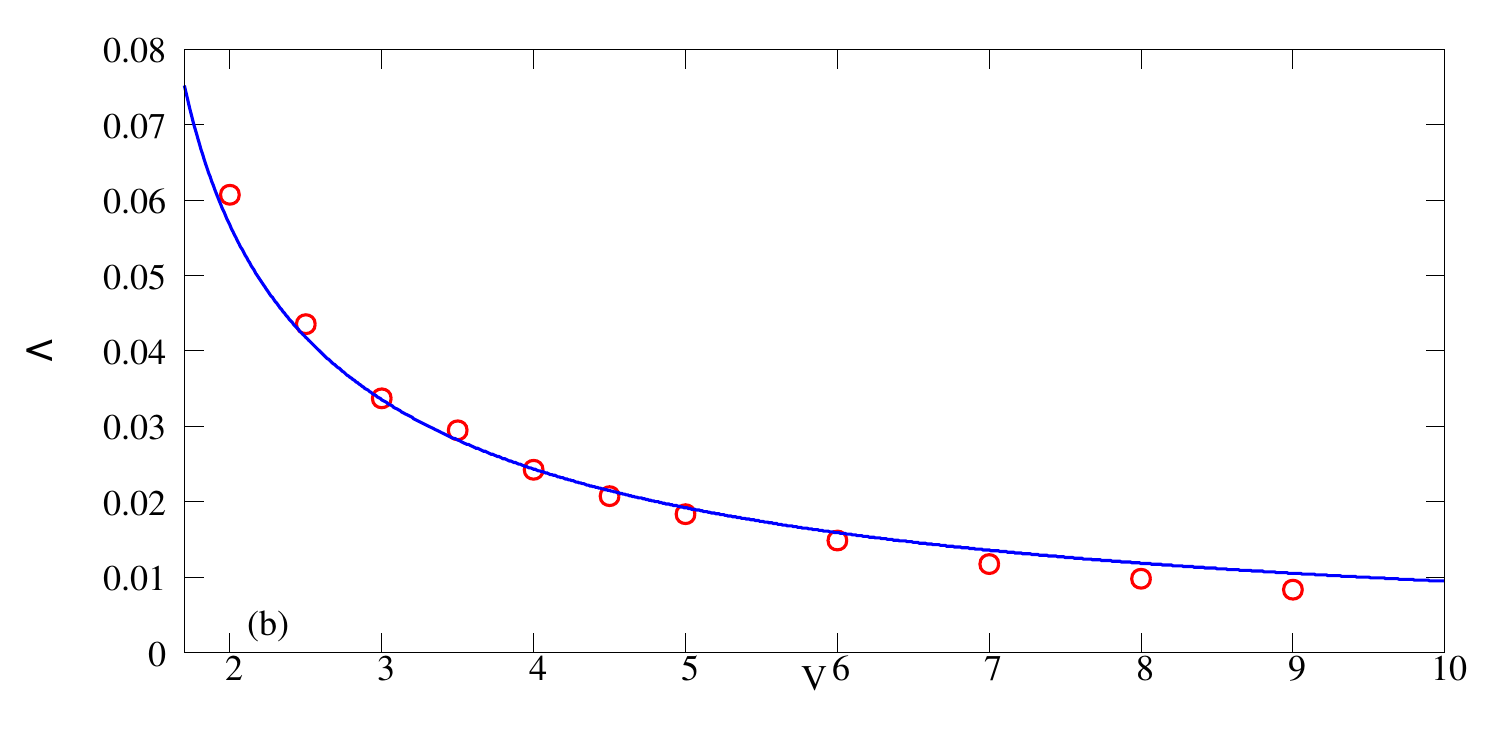 }
\caption{(a) The norm vs coordinate $x$ for selected values of $V$ (markers), together with linear fits. (b) The spatial LE vs $V$ (markers), and the estimate \eqref{eq:forsle},\eqref{eq:splecon} (line).}
\label{fig:ln}
\end{figure}

According to the general theory of Lyapunov vectors in distributed systems~\cite{Pikovsky-Politi-98,Pikovsky-Politi-16}, linear fields  $\tilde U(x,t)$ are highly intermittent, and their logarithms $\log \tilde U(x,t)$ constitute a roughening interface, the mean velocity of which is the Lyapunov exponent. Thus, we calculate the space-dependent norm of the perturbation by averaging  $\log \tilde U(x,t)$ over a large interval $T$
\begin{equation}
N(x)=\frac{1}{T}\int_{t_0}^{t_0+T} \log \tilde U(x,t)\;dt\;.
\label{eq:norm}
\end{equation}
For large $x$, the norm grows linearly with $x$ (Fig.~\ref{fig:ln}(a)), and fitting these linear dependencies one finds the spatial LE (Fig.~\ref{fig:ln}(b)).

One can estimate the spatial LE also from the convective LEs. Consider a perturbation $U(x,t)$ at some position $x$ and time $t$. To this perturbation contribute different components initiated at the boundary $x=0$. We estimate the norm of $U(x,t)$ as the maximum over different contributions. A perturbation at the boundary at time $t-\tau$ contributes to  $U(x,t)$ with the exponential factor $\exp[\lambda(v)\tau]=\exp[\lambda(v)v^{-1}x]$, where $v=x/\tau$ is the corresponding velocity. Thus, we have to maximize $\lambda(v)/v$. This condition yields the relation $\lambda'(v^*)v^*-\lambda(v^*)=0$ for the optimal velocity. Thus, the convective exponent-based estimation of the spatial exponent is 
\begin{equation}
\Lambda=\max_{v}\frac{\lambda(v)}{v}=\frac{\lambda(v^*)}{v^*},\qquad \frac{d\lambda(v^*)}{dv}v^*-\lambda(v^*)=0\;.
\label{eq:forsle}
\end{equation}
Geometrically, this is the slope of the tangential to the profile of the convective LE, as shown for model \eqref{eq:ca} in Fig.~\ref{fig:convleca} with a dashed line. For this model, using the expression for the convective LE in the caption of Fig.~\ref{fig:convleca}, we find $\Lambda\approx 0.31$, which corresponds well with the direct numerical evaluation using \eqref{eq:sple}, which 
yields $\Lambda\approx 0.314$.

By comparing \eqref{eq:forsle} with  expression \eqref{eq:expconvinv}, one can see that 
\[\mu^*=\frac{d\lambda(v^*)}{dv}=\frac{\lambda(v^*)}{v^*},\qquad \mathcal{L}(\mu^*)=0\;.
\]
Thus, $\Lambda=\mu^*$, so
in the terms of the chronotopic LE $\mathcal{L}(\mu)$, the spatial LE is the root 
\begin{equation}
\mathcal{L}(\Lambda)=0\;.
\label{eq:splecon}
\end{equation}
This appears to be quite natural, as the spatial exponent is calculated for a stationary on average temporal field, so it corresponds to the value of $\mu$ at which the corresponding temporal chronotopic exponent $\mathcal{L}$ vanishes.
We compare this expression with the numerics for spatial LE of the NHKS model in Fig.~\ref{fig:ln}(b).

Positive spatial LE explains observation of a non-perfect synchrony in the simulations of the NHKS equation Fig.~\ref{fig:syn}(b). The reason lies in small perturbations at the level of accuracy of double-precision number representation in numerical simulations. These perturbations are unavoidable because the coupling of different sites in the discrete representation \eqref{eq:ksdiscr} is bidirectional. Estimating the level of these perturbations as $10^{-15}$, and taking into account that for $V=2$  the spatial LE is $\approx 0.06$, we conclude that perturbations reach level one at distance $\approx \ln(10^{15})/0.06\approx 575$, what corresponds to the right border of the synchronized domain in Fig.~\ref{fig:syn}(b). Exactly in the same way, positive spatial LE also explains an exponential growth of stochastic turbulence component in experiments, as is shown in Fig. (27) of Ref.~\cite{borodulin2014properties}.

\begin{figure}[!htb]
\centering
\includegraphics[width=\columnwidth]{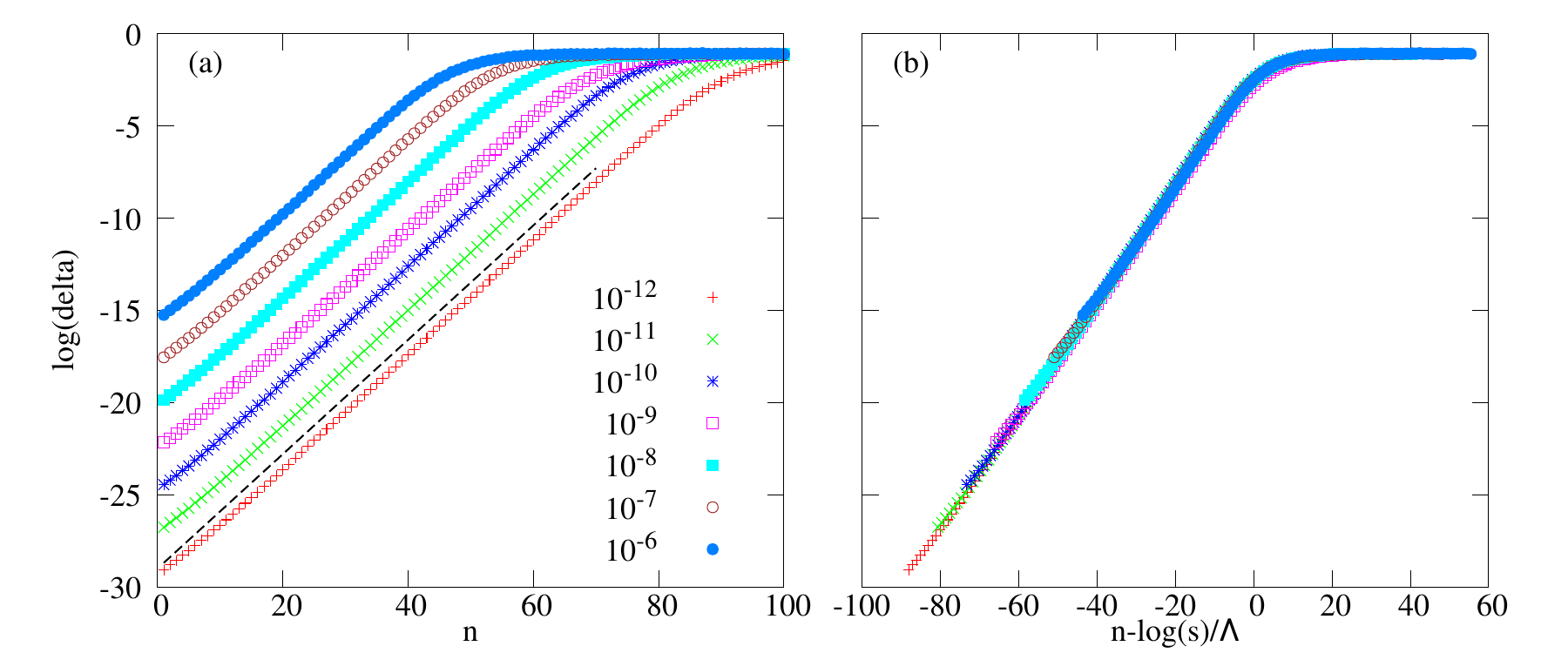}
\caption{(a) Discrepancy between two identical systems \eqref{eq:ca}, driven by the same turbulent signal $f(t)$; to one of the systems, a small additional Gaussian noise with standard deviations as in the markers keys is added. The dashed line has slope $\Lambda=0.314$.  (b) The same data as in panel (a), but shown vs the rescaled distance $n-\frac{\ln\delta(0)}{\Lambda}$. The norm \eqref{eq:norm} is used for $\delta$.}
\label{fig:scal}
\end{figure}

To establish scaling properties of the growth of the stochastic component, let us denote its level along the system as $\delta(X)$, with a fixed inlet level $\delta(0)$. If $\delta(0)$ is small, growth of the stochastic component follows the exponential law \eqref{eq:sple}, i.e. 
\begin{equation}
\delta(X)\sim \delta(0)\exp[\Lambda X]=\exp\left[\Lambda\left(X-\frac{\ln\delta(0)}{\Lambda}\right)\right], 
\label{eq:scal}
\end{equation}
until a saturation at the level $\delta\approx 1$ occurs. Scaling expression \eqref{eq:scal} shows, that in the setup where the same input signal (up to some noisy discrepancy $\delta(0)$) is applied to two identical replicas of a convectively unstable medium, then the evolution of the difference between two systems downstream can be expressed as an function of single quantity $X-\Lambda^{-1}\ln\delta(0)$. We illustrate this in Fig.~\ref{fig:scal}, where we use model \eqref{eq:ca} because it allows for a perfect control of the inlet perturbations.

We should mention, however, that the spatial LE provides, for a setup like depicted in Fig.~\ref{fig:syn}(b), only an estimate of the distance at which the reliability breaks due to uncontrolled perturbations at the boundary and in the bulk. The reason is that the spatial LE is well-defined only for a spatially stationary regime like one corresponding to the bottom panel of Fig.~\ref{fig:nstks}. Closer to the inlet, there is a domain where statistical properties of the field are not stationary but heavily depend on the driving force $f(t)$. Correspondingly, spatial amplification of perturbations in this non-stationary domain may deviate from that described by the spatial LE.  

\section{Periodic and quasiperiodic input}
\label{sec:reg}

Because the considered systems \eqref{eq:ks1} and \eqref{eq:ca} are stable in the sense of negative usual LEs, their response to a periodic input signal $f(t)=f(t+T)$ is periodic. This is mostly transparent for the chain of ODEs \eqref{eq:ca}. Indeed, each element is a linear filter driven by a signal from the left neighbor, and if the driving signal is periodic, the output, after a long enough transient, will also be periodic $u_n(t)=u_n(t+T)$. For the NHKS equation, the temporal periodicity of the field is confirmed by numerics. We, however, postpone detailed discussion of this regime to Section~\ref{sec:std}.

Here, we focus on nontrivial properties of the quasiperiodic driving, where, in the simplest case, the force contains two incommensurate frequencies
\begin{equation}
f(t)=a_1\cos(\omega_1t+\phi_1)+a_2\cos(\omega_2 t+\phi_2)\;.
\label{eq:qpf}
\end{equation}
For definiteness, in the simulation below, we adopt the golden mean ratio of frequencies 
\begin{equation}
\frac{\omega_2}{\omega_1}=\frac{\sqrt{5}-1}{2}=\frac{T_1}{T_2}\;,
\label{eq:gm}
\end{equation}
where $T_{1,2}=2\pi/\omega_{1,2}$ are two incommensurate periods.

The arguments similar to those used for a periodic forcing indicate that the field remains quasiperiodic in time at all distances from the inlet. Again, this is evident for the chain \eqref{eq:ca} because each unit is a quasiperiodically driven linear filter. In a more general context, one should exclude the possibility of strange nonchaotic regimes which may appear in weakly stable systems under quasiperiodic forcing~\cite{Feudel-Kuznetsov-Pikovsky-06}. According to the results of Ref.~\cite{Stark-97}, for strong enough stability, quasiperiodicity of the output is ensured. Applied to the NHKS equation \eqref{eq:ks1}, this means that the usual temporal LE should be negative enough, i.e., parameter $V$ should be large enough.

Although the regime in the quasiperiodically driven convective system remains quasiperiodic, 
it becomes more and more complex downstream, and at large distances, it is practically indistinguishable from a random signal. This has been demonstrated for system \eqref{eq:ca} in \cite{Pikovsky-88,Pikovsky-89b}. Below, we show this spatial development of field complexity for the NHKS model \eqref{eq:ks1}.

We explore below two characterizations of the quasiperiodic dynamics. In the first one (Section~\ref{sec:tps}), we consider a discrete phase-space representation of the motion. In the second approach (Section~\ref{sec:tsp}), we consider the spectrum of the process.
\subsection{Evolution of a torus in the phase space}
\label{sec:tps}

\begin{figure}[!htb]
\centering
\includegraphics[width=\columnwidth]{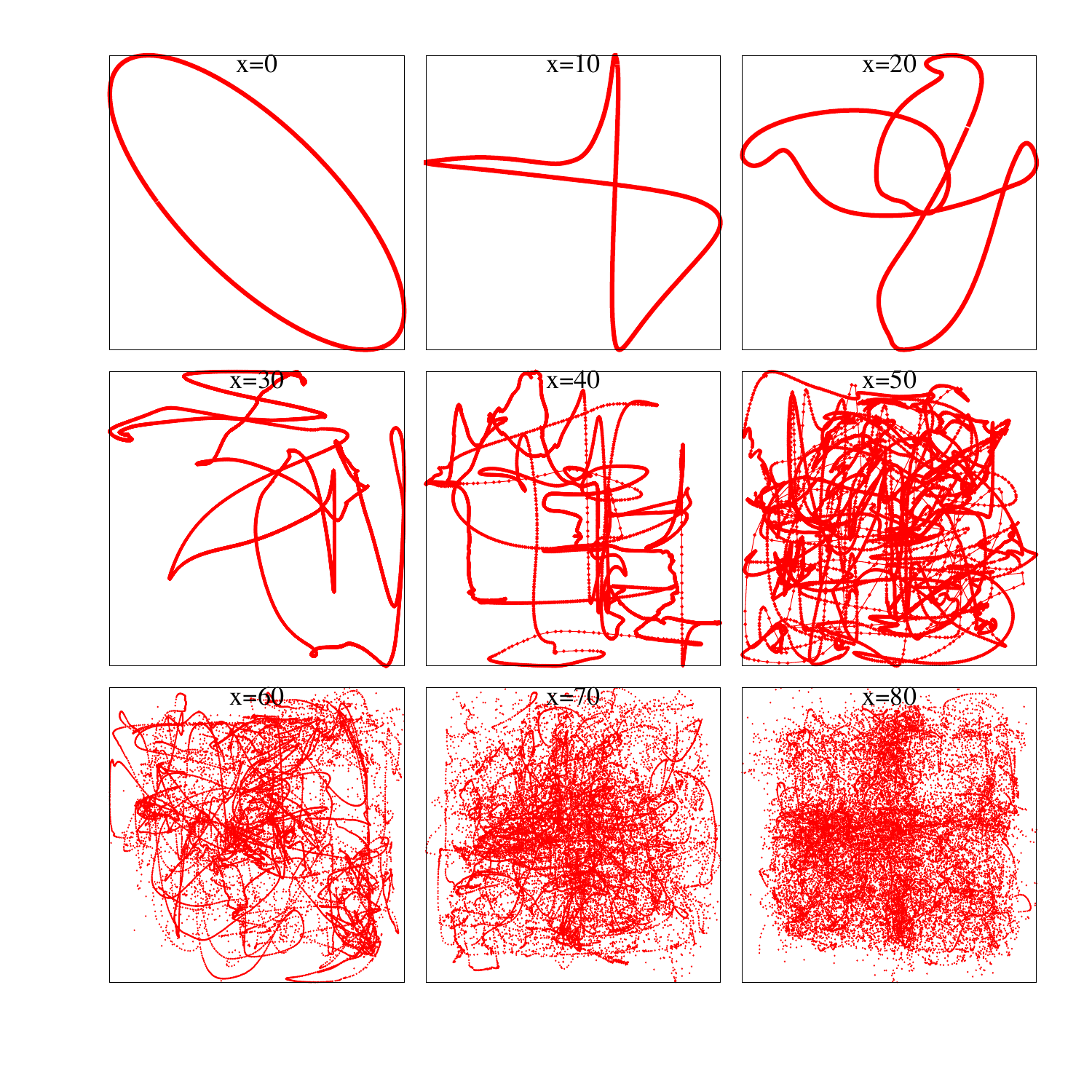}
\caption{ Images of the torus at different $x$ for $V=2$ and $T_1=10$.  For $x=0,10,20$ only the line connecting the points is shown (because the points are extremely dense). For $x=30,40,50$ the points and the line connecting them are shown. One cannot distinguish points for $x=30$, but for $x=40,50$ one can see them because the separation between the points becomes large. For $x=60,70,80$ only points are shown; moreover, for better visibility, the point size is reduced.}
\label{fig:qp2}
\end{figure}

The image of a two-frequency quasiperiodic dynamics is a two-dimensional torus in the phase space. Using a discrete stroboscopic map, one reduces this torus to a one-dimensional closed curve. It is convenient to use one of the basic incommensurate periods for the stroboscopic mapping. Thus, we drive NHKS system \eqref{eq:ks1} with force \eqref{eq:qpf}, and consider the field $u(x,t)$ at discrete times $t=n\frac{2\pi}{\omega_1}=nT_1$. 

We represent the resulting discrete sequence $u_n(x)=u(x,nT_1)$ on the plane $(u_n,u_{n+1})$. Figure \ref{fig:qp2} shows the images at different values of $x$. At $x=0$, the field $u(0,t)$ is the signal  \eqref{eq:qpf}, and correspondingly the image in the stroboscopic map is an ellipse. One can see that this ellipse becomes deformed and folded at larger $x$. The most important feature is that the total length of the closed invariant curve grows with $x$. This means that one needs more points in a time series to fill the curve without visible gaps. In Fig.~\ref{fig:qp2}  we use 
28657 points (this is a Fibonacci number; such lengths of the time series ensure optimal coverage of the invariant curve for the golden ratio of the frequencies). While at $x=30$, the points are dense and form a curve without visible gaps (meaning that gaps are smaller than the dot size in the figure), at $x=40$, one can clearly see regions where the distances between the points are large. For $x=40,50$ we connect the points with a line to visualize the closed invariant curve, but for $x\geq 60$ this is impossible (although at $x=60,70$ one can still see fragments where the points are close to each other and form pieces of the invariant curve). At $x=80$ the picture is practically a random set of points; the underlying line structure is hardly visible.   

\begin{figure}[!htb]
\centering
\includegraphics[width=\columnwidth]{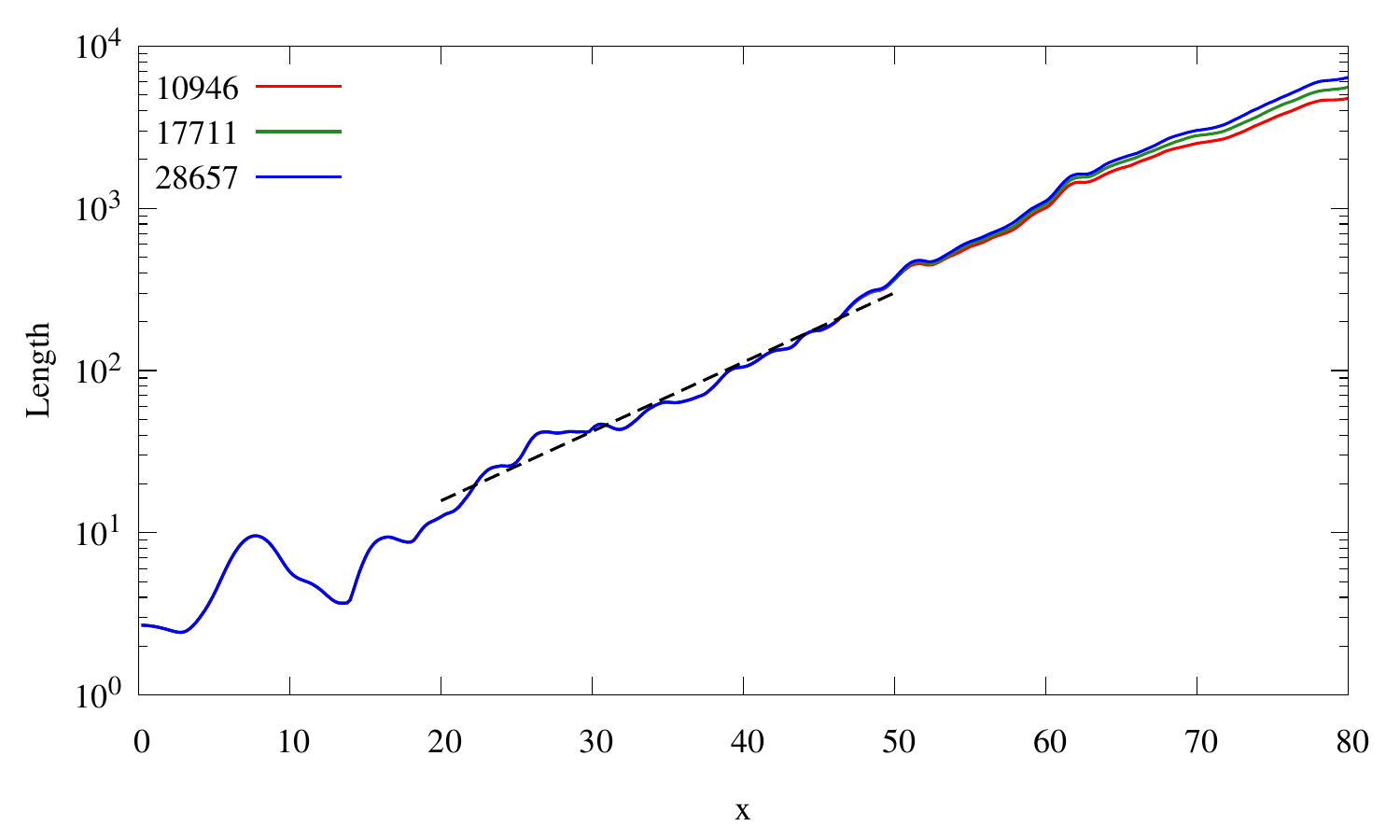}
\caption{ Length of the torus in images like in Fig~\ref{fig:qp2}  vs distance $x$.
One can see that beyond $x=50$ the curves for different Fibonacci numbers of the number of points do not overlap, which corresponds to a bad approximation of the torus as a line for these lengths.}
\label{fig:qp2length}
\end{figure}

We quantify this phenomenon of growing length of the invariant curve in Fig.~\ref{fig:qp2length}. We present here three calculations of the length with different Fibonacci numbers of points. Reliable lengths are those where these curves coincide ($x<50$), which corresponds to the images of Fig.~\ref{fig:qp2}. Fig.~\ref{fig:qp2length} demonstrates that after some oscillations at small $x$, the length grows as $L\propto \exp[\gamma x]$ (black dashed line in Fig.\ref{fig:qp2length}), with $\gamma\approx 0.1$. This exponent characterizes the spatial growth of the signal's complexity and its evolution toward a random process.

\subsection{Evolution of the spectrum}
\label{sec:tsp}

\begin{figure}[!htb]
\centering
\includegraphics[width=\columnwidth]{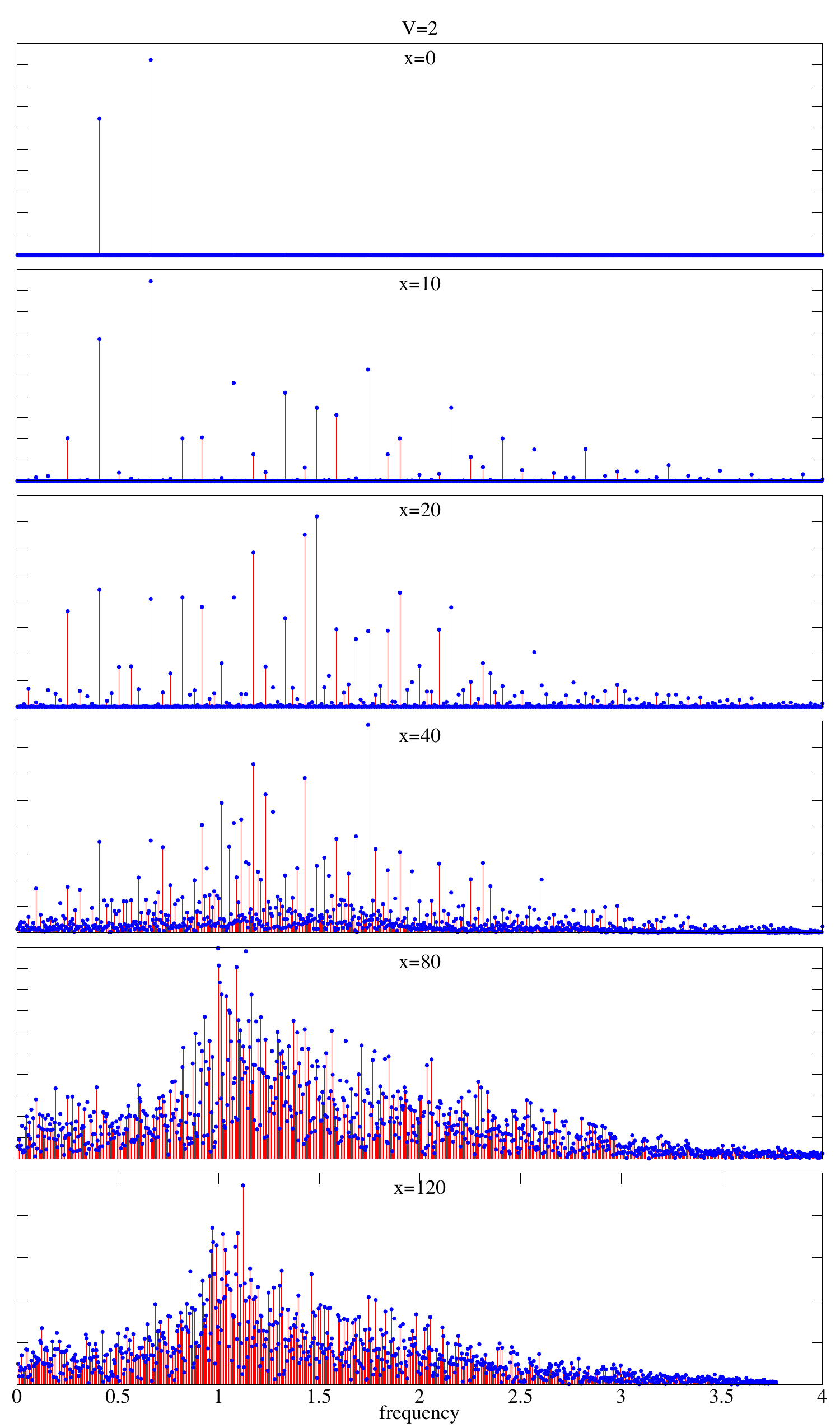}
\caption{Spectra at different positions for $T_1=10$. The vertical scales at different $x$ are different. At $x=0$ only two harmonics are present; the amplitudes of the combinational harmonics grow with $x$. The spectra at $x=80$ and $x=120$ are statistically equivalent.}
\label{fig:qpsp}
\end{figure}

The driving force \eqref{eq:qpf} has just two spectral components at two incommensurate frequencies $\omega_1,\omega_2$. A general two-frequency quasiperiodic signal has, generally, non-zero components at all possible combinational frequencies $k_1\omega_1+k_2\omega_2$ with integers $k_1,k_2$, but for a simple smooth torus, the amplitudes of the components with large $|k_1|,|k_2|$ are very small. We show in Fig.~\ref{fig:qpsp} that the combinational harmonics grow with distance $x$ and eventually the spectrum looks like a continuous spectrum. To reliably calculate spectral components, we approximate the signal  \eqref{eq:qpf} with a periodic one, using successive Fibonacci numbers $q_m,q_{m+1}$: $\omega_2=\frac{q_m}{q_{m+1}}\omega_1$. This makes the whole driving periodic with a large period $2\pi q_{m+1}/\omega_1$, and the spectrum contains only harmonics of this small basic frequency $\omega_0=\omega_1/q_{m+1}$. We denote the amplitude of the component with frequency $k\omega_0$ as $W_k$. With increasing $m$, one gets a better and better approximation of the genuine spectrum. Note that because the field $u(x,t)$ is smooth, spectral components with very high frequencies have small amplitudes. Thus, in the calculation of the spectrum, we restricted the range to $k\leq 6 q_{m+1}$.

\begin{figure}[!htb]
\centering
\includegraphics[width=\columnwidth]{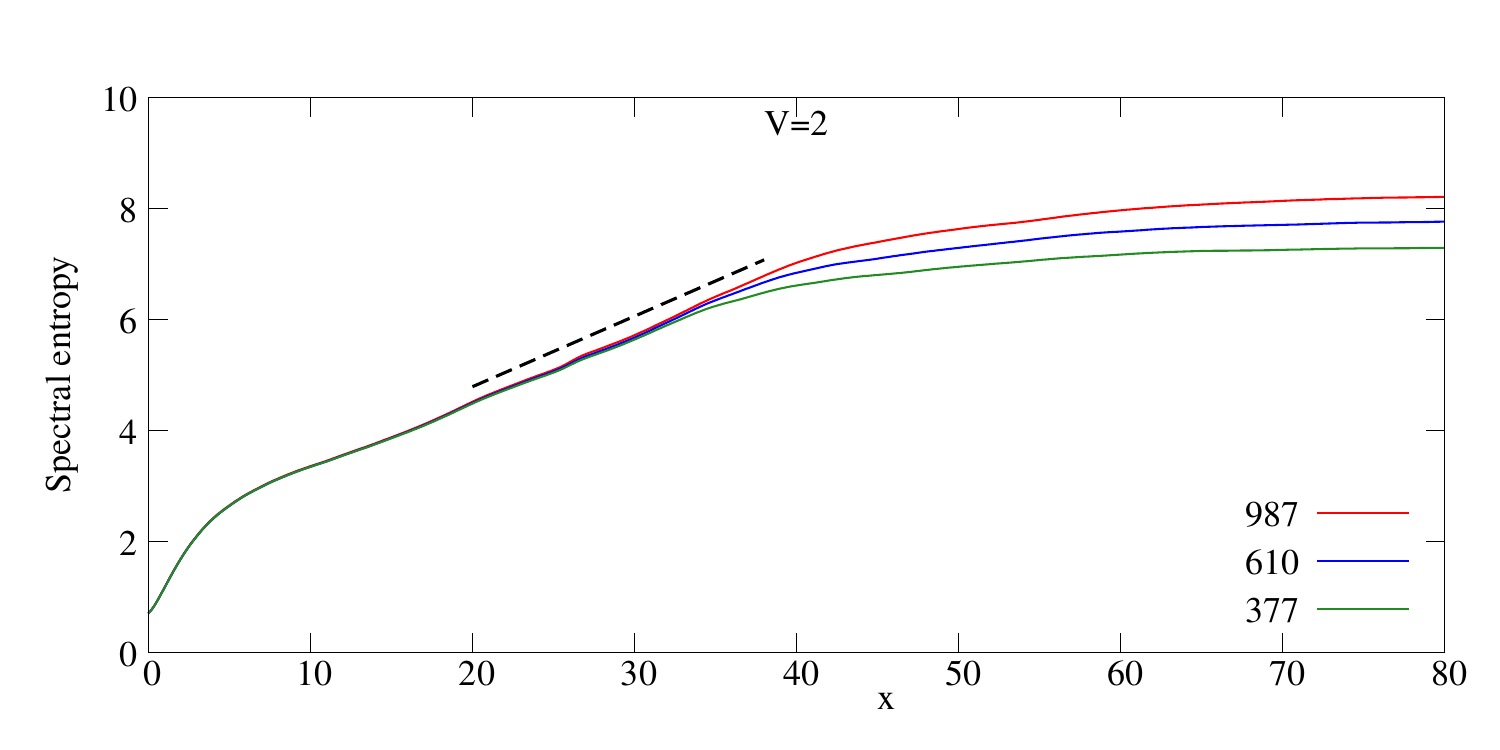}
\caption{Evolution of spectral entropy for different values of $q_{m+1}$. The black dashed line has slope $0.127$.}
\label{fig:qpspen}
\end{figure}

Generation of more and more harmonics can be characterized quantitatively via spectral entropy. 
We define it as 
\[
\mathcal{S}(x)=-\sum_{k=1} w_k(x)\log w_k(x) \;,
\]
where $w_k$ are normalized spectral components
$
w_k=\frac{W_k}{\sum_k W_k}
$.
The evolution of the spectral entropy for several periodic approximations $\omega_2/\omega_1=q_m/q_{m+1}$ where $q_{m}$ are Fibonacci numbers, is presented in Fig.~\ref{fig:qpspen}. One can see three stages: (i) initially for $x\lesssim 5$ there is a strong growth of higher harmonics of the two initial frequencies; (ii) for $x\gtrsim 5$ there is a linear growth of entropy until a saturation, which depends on the total number of harmonics ($\mathcal{S}_{max}\propto \ln q_{m+1}$). Because the entropy is a logarithmic measure of complexity, its logarithmic growth corresponds to the exponential growth of the invariant curve length in Section \ref{sec:tps}. Both figures \ref{fig:qp2length} and \ref{fig:qpspen} illustrate exponential spatial growth of complexity at a quasiperiodic driving, so that at large distances, with a finite resolution, one cannot distinguish a complex quasiperiodic process from a random one.

\section{Space-time duality}
\label{sec:std}

In the convective space-time chaos, one considers the dynamics in the domain $x\geq 0,\;t\geq 0$, subject to initial conditions at $t=0$ and boundary conditions at $x=0$. The perturbations propagate with some velocity from both boundaries into the bulk of the domain, as illustrated in Fig.~\ref{fig:ksfield}. Due to the symmetry $t\leftrightarrow x$ of the setup, one can exchange space and time: one interprets $x$ as ``time''  and $t$ as ``space''. Below, we will keep this terminology: we will use ``time'' and ``space'' in quotation marks, if we want to denote the variable along which the evolution takes place, and the variable in which the corresponding ``initial conditions'' are set, respectively. 

For the discrete model of a chain of amplifiers~\eqref{eq:ca}, one interprets $n$ as discrete ``time'', so that the evolution in $n$ can be written as
\begin{equation}
u_{n+1}(t)=\hat{D} f(u_n(t))\;.
\label{eq:spamap}
\end{equation}
Here $\hat{D}$ is the ``coupling operator'', a spectral representation of which is  $\hat{D}(\omega)=(1+i\omega)^{-1}$. Such a discrete in ``time'' and continuous in ``space'' model has been used in~\cite{Kuznetsov-Pikovsky-86} (see \cite{Kuznetsov-82} for application of the model~\eqref{eq:spamap} to time-delay systems).

Representation \eqref{eq:spamap} is mostly suitable for studying periodic in ``space'' $t$ fields, as such a setup is typical in studies of space-time chaos. This problem corresponds exactly to the problem of spatial evolution under periodic forcing, formulated at the beginning of Section~\ref{sec:reg}. The period of the force $T$ corresponds to the period in ``space'' in \eqref{eq:spamap}. The usual largest LE of  \eqref{eq:spamap} determines whether the evolution in ``time'' is chaotic or not, for large $T$ it is the same as the spatial LE \eqref{eq:sple}. With this space-time duality, reliability discussed in Section~\ref{sec:rel} corresponds to the fact, that the ``temporal'' evolution from two identical ``initial conditions'' 
will be exactly the same. If, however, the ``initial conditions'' are slightly non-identical, then the discrepancy grows exponentially and after a logarithmically short ``time''  the states of the two runs will be different (onset of stochastic turbulence). Furthermore, one can characterize the instability related to a non-perfect periodicity of the driving with the Bloch-Lyapunov exponent~\cite{Pikovsky-88,Pikovsky-Politi-16}, but we will not go into details.

Let us now discuss space-time duality for the NHKS equation \eqref{eq:ks1}. In this equation, the variables $x$ and $t$ enter highly asymmetrically, and the first derivative with respect to $t$ indicates that evolution is along this variable. However, one can argue that for large velocities $V$, one can exchange $t$ and $x$ in an approximate way.
Let us first make a change of variables $z=Vt$. Then \eqref{eq:ks1} reduces to
\begin{equation}
\partial_z u+\partial_x u+\frac{1}{V}(u\partial_x u+\partial_{xx}u+\partial_{xxxx}u)=0 \;.
\label{eq:ks-t1}
 \end{equation}
 If we assume $V$ to be large and neglect terms $\sim 1/V$, then in the leading order $\partial_z=-\partial_x$. Then, we substitute   higher derivatives in $x$ by higher derivatives in $z$ to obtain
 \[
 \partial_z u+\partial_x u+\frac{1}{V}(-u\partial_z u+\partial_{zz}u+\partial_{zzzz}u)=0 \;.
 \]
Now we change the variable $x=V\tau$ and obtain
\begin{equation}
\partial_\tau u+V\partial_z u-u\partial_z u+\partial_{zz}u+\partial_{zzzz}u=0\;.
\label{eq:ks-t2}
\end{equation}
This is the original NHKS (up to transformation $u\to -u$), where the space and time are exchanged: the ``time'' variable $\tau$ corresponds to $x$ and the ``space'' variable $z$ corresponds to $t$. Such a
 duality is mostly simple for linear equations. Assume that one has a dispersion relation $R(\omega,k)=0$. Suppose one wants to follow the evolution of a spatial profile over time. In that case, it is natural to
  assume that the wavenumber $k$ is real, and to express the frequency $\omega$, which then will be generally complex, as a function (typically a polynomial) of $k$. This leads to a PDE in the form
   \eqref{eq:ks1}. If, in contradistinction, one is interested in the evolution in space of the inlet
    field depending on time, then it is natural to assume the frequency $\omega$ to be real, and to
     express the complex wavenumber $k$ as a function (typically a polynomial) of frequency $\omega$; this leads to an equation in the form \eqref{eq:ks-t2}. We note here that periodic external forcing
      discussed at the beginning of Section~\ref{sec:reg} corresponds to periodic in ``space'' boundary conditions in the dual representation \eqref{eq:ks-t2}. Remarkably, an equation dual to the
       Korteveg - de Vries equation has been recently discussed in Ref.~\cite{smilga2021exactly}, however, without consideration of the corresponding duality of initial and boundary conditions.

Similar to the discrete case~\eqref{eq:spamap}, the spatial LE in the original \eqref{eq:ks1} corresponds to the ``temporal'' LE in \eqref{eq:ks-t2}. As we have performed the transformation $z=Vt$, the spatial exponent is $\Lambda=\lambda_{max}V^{-1}$; this relation corresponds to what follows from \eqref{eq:forsle} for the NHKS equation at large $V$.

For the problem of waves on an inclined fluid flow, for which the NHKS equation \eqref{eq:ks1} has been derived, the dual representation \eqref{eq:ks-t2} makes sense for large values of parameter $V$. There is, however, a situation where a natural way to describe the dynamics is to consider the evolution along the spatial variable and treat time dependence as effective space dependence. This is the case of nonlinear optics in one-dimensional media (fibers)~\cite{turitsyn2016dissipative,Agrawal-13}, where the speed of light is certainly always large. A typical equation of the slowly varying complex amplitude $A(z,t)$ of the radiation field is a generalized nonlinear Schr\"odinger equation
\begin{equation}
\partial_zA=\beta_1\partial_tA-i\frac{\beta_2}{2}\partial_{tt}A+\frac{\beta_3}{6}\partial_{ttt}A+i\gamma|A|^2A+ GA\;.
\label{eq:opt}
\end{equation}
Here, the parameters $\beta_{1,2,3}$ describe the group velocity and the dispersion. The ``time'' coordinate $z$ is in fact the distance along the fiber, and equation \eqref{eq:opt} describes the spatial evolution of the input signal $A(0,t)$. To simulate mode-locked fiber lasers, one uses generalizations of \eqref{eq:opt} that include additional dissipative linear and nonlinear terms~\cite{turitsyn2016dissipative}. The resulting equation is the cubic Ginzburg-Landau equation, which can also demonstrate space-time chaos.

Finally, we mention that in the context of hydrodynamics, the space-time duality is employed in the ``One-Way Navier-Stokes'' approach for calculations of a compressible jet~\cite{towne2015one} and of a boundary layer transition~\cite{sleeman2025boundary}.

\section{Discussion}
\label{sec:concl}

In conclusion, we have demonstrated how the properties of deterministic and stochastic turbulence, observed in experiments~\cite{borodulin2011experimental,borodulin2013experimental,kachanov2013hypothesis,borodulin2014properties}, can be explained through the concept of convective space-time chaos. Let us consider again the main experimental features outlined in the Introduction (Section~\ref{sec:intro}), together with their dynamical interpretation.
\begin{enumerate}
\item Evolution, along the spatial coordinate, of an imposed random signal toward a regime where statistical characteristics of space-time chaos do not depend on the forcing. The most straightforward interpretation of this feature is through the space-time duality (Section~\ref{sec:std}). In the dual representation, the boundary forcing corresponds to ``initial conditions'', and development in space to evolution in ``time''. It is known that at the initial stage, the dynamical system remembers initial conditions. However, after the trajectory converges to an attractor, initial conditions are forgotten, and one observes, in the case of a chaotic attractor, irregular chaotic dynamics, statistical properties of which do not depend on the initial conditions. 
\item Driven dynamical systems with negative Lyapunov exponents demonstrate stable synchronization by external driving (if several identical systems are driven by the same noise), which is equivalent to reliability (if one system is driven several times by the same signal). This explains why, in repeating experiments, the same pattern of fluctuations is observed. Stability here is understood as stability toward variation of initial conditions, i.e. in the sense of the usual largest Lyapunov exponent, which is negative for convective space-time chaos in confined geometries without (or with very small) reflections.
\item Although reliability/synchrony holds for identical systems and identical driving, this property is sensitive to violations of identity. This sensitivity to the forcing, which can be referred to as ``sensitive dependence on boundary conditions'', is quantified by the spatial Lyapunov exponent, which is positive for convective space-time chaos. If one applies the space-time duality, then the spatial Lyapunov exponent becomes the usual temporal Lyapunov exponent, and sensitivity to boundary conditions becomes the usual sensitive dependence on initial conditions in chaos. This explains the transition from the deterministic to the stochastic turbulence in experiments~\cite{borodulin2011experimental,borodulin2013experimental,kachanov2013hypothesis,borodulin2014properties}. The size of the domain where deterministic turbulence is observed, is roughly proportional to the logarithm of the deviations from identity, and inversely proportional to the spatial Lyapunov exponent (with a possible correction that this exponent is different within transients described in feature 1).
\end{enumerate}

Furthermore, we discussed the transformation of periodic and quasiperiodic signals in convective space-time chaos. Periodic boundary forcing leads to a periodic in time field downstream. If the period is large, much larger than the characteristic time scale of temporal correlations of chaos, then at any distance from the inlet, the field will have the same period, but within this period will look like a random one. This, in particular, has been observed in early experiments with the boundary layers~\cite{borodulin2011experimental}, where periodic driving with a period of about 20 characteristic time scales was used. In terms of the space-time duality, this setup corresponds to periodic in ``space'' initial conditions, which remain periodic. A destruction of periodicity in the presence of small non-periodic perturbations is governed by the Bloch-Lyapunov exponent~\cite{Pikovsky-Politi-16}. 

A nontrivial transition from regularity to effective chaos, even in the absence of secondary perturbations,  is observed for a quasiperiodic driving. We provided a twofold characterization of the spatial growth of complexity of a quasiperiodic signal: (i) by virtue of the growth rate of the torus length in the phase space, and (ii) by virtue of the growth of the spectral entropy of the spectral representation of the signal. Quasiperiodic driving appears to be feasible in experiments of type~\cite{borodulin2011experimental,borodulin2013experimental,kachanov2013hypothesis,borodulin2014properties}; for the experiments with a chain of amplifiers see~\cite{Pikovsky-89b}.

The concept of convective space-time chaos generalizes notions of linear absolute and convective instability (in discrete situations, one often speaks about normal and non-normal matrices~\cite{Trefethen-Embree-20}) to secondary instabilities on top of space-time chaos. We note here that the conditions of convectivity for linear and secondary perturbations may differ. Thus, it would be interesting to explore cases where, e.g., linear instability is absolute but secondary instability is convective. One field of application of these concepts is chemical reactions in flow systems~\cite{Kuznetsov-Mosekilde-Dewel-Borckmans-97}. Another interesting question for future studies is, what happens close to the transition from convective to absolute instability, e.g., if parameter $V$ in Eq,~\eqref{eq:ks1} is decreased toward the critical value $V\approx v_0=1.4$. Such a transition also occurs in a Coutte-Taylor flow between rotating cylinders with an additional throughflow, if the throughflow velocity is varied~\cite{tsameret1991convective,babcock1991noise}.

While our approach is inspired by hydrodynamic experiments, the results are not restricted to this realm, but are applicable to other cases of spatial development of chaos and turbulence, e.g., in plasma physics and electronics. Our study once again stresses the well-established importance of the inlet perturbations for the evolution of convective instabilities. 

In some cases, one is not interested in this evolution, and applies the inlet perturbations mimicking a fully developed regime (see review~\cite{wu2017inflow} for such an approach in computational fluid dynamics). In such a setup the feature 1 is not applicable, but still, the reliability property can be followed by repeating the exact form of the inflow turbulence in several numerical experiments and comparing the outputs. Our results of Sec.~\ref{sec:sple} imply that in numerical experiments, even small uncontrollable round-off errors will result in a loss of reliability (repeatability) in the output. Moreover, the spatial evolution of violations of reliability allows for the estimation of the spatial Lyapunov exponent. 

Another often implemented option is to apply a small periodic inlet perturbation to otherwise ideal flow; in the boundary layer studies, this has been done in experiments~\cite{kachanov1984resonant,cheng2021h} and in numerical simulations~\cite{rist1995direct,sayadi2013direct,lin2025boundary}. As discussed above, formally speaking, a periodic inlet perturbation in a convectively unstable medium results in a purely periodic response. However, such a regime is typically unstable, and this instability can be described with the Bloch-Lyapunov exponent, which generalizes the usual secondary instability approach~\cite{herbert1988secondary} to the case of irregular in space variations of a temporally periodic field. Thus, the break of periodicity (cf. recent discussion of this phenomenon in \cite{lin2025boundary}) is similar to the break of reliability. The distance at which this happens depends on the Bloch-Lyapunov exponent and the level of irregular perturbations (in the case of a boundary layer, these perturbations are discussed in Ref.~\cite{saric2002boundary}). Our numerical simulations of the PDE model suggest that it is extremely difficult, if altogether possible, to reduce these irregular perturbations to zero. However, it is always possible to introduce them in a controlled way at a level higher than uncontrollable round-off errors, like we did in Section~\ref{sec:sple}.

%
%

\begin{acknowledgements}
The author thanks M. Zaks and A. Nepomnyashchy for useful discussions. This study was inspired by the talk of E. Bodenschatz at the Dynamics Days Europe 2025 (Thessaloniki, June 2025).
\end{acknowledgements}








\end{document}